\documentclass[a4paper,12pt]{article}

\usepackage{amssymb,amsmath,bm}
\usepackage{a4wide}
\usepackage{color}
\usepackage{slashed}
\usepackage{graphicx}
\usepackage[latin1]{inputenc}
\usepackage{amsfonts}
\usepackage{lscape}
\usepackage{amsthm}
\usepackage{booktabs}
\usepackage{multirow,geometry,mathrsfs}
\usepackage{array}
\usepackage{cancel}
\usepackage{ulem}
\usepackage{rotating}
\usepackage[numbers, sort]{natbib}
\usepackage{float}
\usepackage{relsize,exscale}
\usepackage[toc,page]{appendix}
\usepackage{multicol}
\usepackage{color}
\usepackage{hyperref}
\definecolor{azur}{rgb}{0,0.498,1}
\definecolor{caeruleum}{rgb}{0.208,0.478,0.718}
\definecolor{indigo_elec}{rgb}{0.435,0,1}
\definecolor{vert}{rgb}{0,0.6,0.2}

\usepackage{eurosym}
\usepackage{feynmp}
\usepackage[small,bf]{caption}
\newcommand{\be}{\begin{equation}}
\newcommand{\ee}{\end{equation}}

\begin{document}
\def\slashq{q\kern -.450em {/}}
\newcommand{\beqn}{\begin{eqnarray}}
\newcommand{\eeqn}{\end{eqnarray}}
\newcommand{\bea}{\begin{eqnarray}}
\newcommand{\ena}{\end{eqnarray}}
\newcommand{\ra}{\rightarrow}
\def\neuto{\tilde\chi^0_1}
\def\neuti{\tilde\chi^0_i}
\def\neutt{\tilde\chi^0_2}
\def\neuth{\tilde\chi^0_3}
\def\neutf{\tilde\chi^0_4}
\def\neutfi{\tilde\chi^0_5}
\def\chargi{\tilde\chi^+_i}
\def\charg{\tilde\chi^+_1}
\def\chargt{\tilde\chi^+_2}
\def\chargm{\tilde\chi^-_1}
\def\chargtm{\tilde\chi^-_2}
\def\stopl{\tilde t_1}
\def\stoph{\tilde t_2}
\def\sbotl{\tilde b_1}
\def\sboth{\tilde b_2}
\def\mneut{m_{\tilde{\chi}^0_1}}
\def\mchi{m_{\tilde{\chi}^0_i}}
\def\mneutt{m_{\tilde{\chi}^0_2}}
\def\mneuth{m_{\tilde{\chi}^0_3}}
\def\mneutf{m_{\tilde{\chi}^0_4}}
\def\mchar{m_{\tilde{\chi}^+_1}}
\def\mchart{m_{\tilde{\chi}^+_2}}
\def\msel{m_{\tilde{e}_L}}
\def\mser{m_{\tilde{e}_R}}
\def\mslo{m_{\tilde{\tau}_1}}
\def\mslt{m_{\tilde{\tau}_2}}
\def\msul{m_{\tilde{u}_L}}
\def\msur{m_{\tilde{u}_R}}
\def\msdl{m_{\tilde{d}_L}}
\def\msdr{m_{\tilde{d}_R}}
\def\msto{m_{\tilde{t}_1}}
\def\mstt{m_{\tilde{t}_2}}
\def\msbo{m_{\tilde{b}_1}}
\def\msbt{m_{\tilde{b}_2}}
\def\drbar{\overline {\textrm{DR}}}
\def\sloops{{\tt{SloopS}}}
\def\tb{t_\beta}
\def\noi{\noindent}

\newcommand{\GC}[1]{{\color{vert} #1}}
\newcommand{\GCd}[1]{\GC{\sout{#1}}}
\newcommand{\gb}{\textcolor{blue}}
\newcommand{\old}{\textcolor{red}}

\begin{titlepage}
\begin{center}
\vspace*{-1cm}
\begin{flushright}
LAPTH-009/16\\
LPSC16022

\end{flushright}

\vspace*{1.6cm}
{\Large\bf One-loop renormalisation of the NMSSM in SloopS : 1. the neutralino-chargino and sfermion sectors.} 

\vspace*{1cm}\renewcommand{\thefootnote}{\fnsymbol{footnote}}

{\large 
G.~B\'elanger$^{1}$\footnote[1]{Email: belanger@lapth.cnrs.fr},
V.~Bizouard$^{1}$\footnote[2]{Email: bizouard@lapth.cnrs.fr},
F.~Boudjema$^{1}$\footnote[3]{Email: boudjema@lapth.cnrs.fr },
G.~Chalons$^{2}$\footnote[4]{Email: chalons@lpsc.in2p3.fr},
} 

\renewcommand{\thefootnote}{\arabic{footnote}}

\vspace*{1cm} 
{\normalsize \it 
$^1\,$LAPTh, Universit\'e Savoie Mont Blanc, CNRS, B.P.110, F-74941 Annecy-le-Vieux Cedex, France\\[2mm]
$^2\,${Laboratoire de Physique Subatomique et de Cosmologie, Universit\'e
Grenoble-Alpes, CNRS/IN2P3, 53 Rue des
Martyrs, 38026 Grenoble, France\\[2mm]
}}

\vspace{1cm}

\end{center}
\begin{abstract}

We have completed the one-loop renormalisation of the Next-to-Minimal Supersymmetric Standard Model (NMSSM) allowing for and comparing between different renormalisation schemes. A special attention is paid to on-shell schemes. We study a
variety of these schemes based on alternative choices of the physical input parameters. In this paper we present our approach to the renormalisation of the NMSSM and report on our results for the neutralino-chargino and sfermion
sectors. We will borrow some results from our study of the Higgs sector whose full discussion is left for a separate publication. We have implemented the set up for all the sectors of the NMSSM within \sloops, a code for the
automatic computation of one-loop corrections initially developed for the standard model and the MSSM. Among the
many applications that allows the code, we present here the one-loop corrections to neutralino masses and to partial widths of neutralinos and charginos into final states with one gauge boson. One-loop electroweak and QCD corrections to the partial widths of third generation sfermions into a fermion and a chargino or a neutralino are also computed.
\end{abstract}

\end{titlepage}

\section{Introduction}
\label{sec:intro}
Supersymmetry has long been considered as the most natural extension of the standard model that can address the
hierarchy problem while providing a dark matter candidate.
The discovery of a Higgs boson with a mass of 125 GeV whose properties are compatible with those of the Standard
Model is a great achievement of the first Run of the LHC
~\cite{Aad:2012tfa,Chatrchyan:2012xdj} and in some sense supports supersymmetry. Indeed, one can argue that a Higgs
with a mass below 130 GeV is a prediction of the minimal supersymmetric standard model (MSSM).
However, the fact that the observed Higgs mass is so close to the largest value that can be achieved in the MSSM, a
value obtained by requiring a rather heavy supersymmetric spectrum, raises the issue of
naturalness~\cite{Barbieri:1987fn,Hall:2011aa}. Another issue with the MSSM is the $\mu$
problem~\cite{Ellwanger:2009dp}. Namely why $\mu$, a supersymmetry preserving mass parameter as it appears in the
superpotential through the operator mixing the two (superfield) Higgs doublets $\mu \hat{H}_d\cdot \hat{H}_u$,
should be, for a viable phenomenology, small {\it i.e.} of the order the electroweak scale, whereas one expects its
value to be rather of order the cut-off scale. Both
these problems are solved in the singlet extension of the MSSM, the
Next-to-Minimal Supersymmetric Standard Model (NMSSM) where the $\mu$ parameter
is generated dynamically through  the vacuum expectation value of the
scalar component of the additional singlet superfield. Moreover, as a bonus
new terms in the superpotential are now present and give a contribution to the quartic Higgs couplings beside the
gauge induced quartic coupling of the MSSM. These new contributions can lead to an increase of the tree-level mass
of the lightest Higgs, thus more easily explaining the observed value of the Higgs
mass~\cite{BasteroGil:2000bw,Ellwanger:2006rm} without relying on very large corrections from the stop/top sector.
Although fine-tuning issues remain~\cite{Ellwanger:2011mu,Gherghetta:2012gb,Ellwanger:2012ke,Binjonaid:2014oga} they are not as severe as in the MSSM.

\noindent The Higgs discovery has thus led to a renewed interest in the NMSSM both at the theoretical and experimental
level with new studies of specific signatures of the NMSSM Higgs sector ~\cite{King:2012tr,King:2014xwa} and/or of
the neutralino and sfermion
sectors~\cite{Dreiner:2012ec,Vasquez:2012hn,Ellwanger:2014hia} being pursued at
the LHC. With the exciting possibility of discovering new particles at the
second Run of the LHC, it becomes even more
important
for a correct interpretation of a future new particle signal to know precisely the particle spectrum as well as to
make precise predictions for the relevant production and decay processes.\\

\noi The importance of loop corrections to the Higgs mass in supersymmetry cannot be
stressed enough. After all, it is because of  radiative corrections that the MSSM has survived. The large radiative
corrections from the top and stop sector
are necessary to raise the Higgs mass beyond the bounds imposed by LEP and to
bring it in the range compatible with the LHC.  
Higher-order corrections are also of relevance for supersymmetric particles, higher-order SUSY-QCD and electroweak
corrections to the full SUSY spectrum have been computed for some time in the MSSM and are incorporated in several
public codes~\cite{Allanach:2001kg,Djouadi:2002ze,Paige:2003mg,Porod:2003um}.
More recently higher-order corrections to Higgs and sparticle masses have been extended to the NMSSM
~\cite{Degrassi:2009yq,Staub:2015aea}.
Several public codes incorporate these corrections with different scopes and approximations,
{\tt NMSSMTools}~\cite{Ellwanger:2005dv,Ellwanger:2006rn}, {\tt SPheno}~\cite{Staub:2010ty,Porod:2011nf},
{\tt SoftSUSY}~\cite{Allanach:2013kza}, {\tt NMSSMCalc}~\cite{Baglio:2013iia} and {\tt FlexibleSUSY}~\cite{Athron:2014yba}. See also the recent work~\cite{ Drechsel:2016jdg} on the corrections to the Higgs masses in the NMSSM. Moreover,
higher-order corrections to decays have also been computed with some of these codes
~\cite{Liebler:2010bi,Das:2011dg,Baglio:2013vya,Baglio:2015noa}.\\

\noi The code {\tt SloopS} was developed for the MSSM with the objective of computing one-loop corrections for
collider and dark matter observables in supersymmetry. The complete renormalisation of the model was performed in
~\cite{Baro:2008bg,Baro:2009gn} and several renormalisation schemes were implemented. This code relies on an
improved version of LanHEP~\cite{Semenov:1998eb,Semenov:2008jy,Semenov:2014rea} for the generation of Feynman rules
and counter terms. The model file generated is then interfaced to {\tt FeynArts}~\cite{Hahn:2000kx}, {\tt
FormCalC}~\cite{Hahn:1998yk} and {\tt LoopTools} for the automatic computation of one-loop
processes~\cite{Hahn:2000jm}.
One-loop corrections to masses, two-body decays and production cross sections at colliders were realized together
with one-loop corrections for various dark matter annihilation
~\cite{Baro:2007em,Baro:2009ip,Baro:2009na,Boudjema:2011ig,Boudjema:2014gza} and coannihilation
processes~\cite{Baro:2009ip}.
{\tt SloopS} has first been extended to include the NMSSM for one-loop processes not requiring renormalisation, such as the rates for gamma-ray lines relevant for Dark Matter
indirect detection \cite{Chalons:2011ia,Chalons:2012xf} and Higgs decays to photons  at
the LHC \cite{Chalons:2012qe,Belanger:2014roa}.\\

\noi The present paper is the first in a serie that describes the implementation of the one-loop corrections for
all sectors of the NMSSM. We will concentrate in this first paper on the details and issues having to do mainly with the
neutralino/chargino sector since the addition of a singlet brings new features compared to the MSSM. We will be brief on the set-up of the renormalisation in the sfermion sector since the particle content is the same as within the MSSM, for this sector we therefore adhere to  the approach given in~\cite{Baro:2009gn} for the MSSM. 
 The chargino-neutralino sector, in particular through the singlet superfield, is quite tied up with the Higgs sector. We will therefore have to borrow some elements from our study of the Higgs sector which we will go over in
more detail in a follow-up paper~\cite{renormalization_higgs}. For the neutralino/chargino sector, different renormalisation schemes are
defined.
In particular we have aimed at studying different on-shell, OS, schemes. The latter are based on choosing a minimal
set of observables, namely masses of physical particles in the NMSSM spectrum to define the set of input parameters
and necessary counterterms which will allow to get rid of all ultra-violet divergences in all calculated
observables. Finding the minimal set of necessary counterterms requires solving a system of coupled equations. For
the case of the NMSSM where mixing between different components occurs and where the same parameters
appears in different sectors, the system of equations can be large. Moreover some choices of the minimal set (and
therefore the relevant coupled equations) will lead to solutions that are extremely sensitive to a particular choice of a parameter which may, in some process, induce large radiative corrections.
It is also possible, when a renormalisation scale $\bar{\mu}$  has been chosen, to follow a simpler implementation of the counterterms, {\it \`a la} $\drbar$, where these
counterterms are {\it pure} divergent terms. In some instances these can also lead to splitting a large system of
coupled equations to a smaller and more manageable system of equations. The renormalisation of the ubiquitous
$t_\beta$ which, at tree-level, represents the ratio of the vacuum expectation values, vev, of the 2 Higgs doublets
is a case in point. We will also study mixed schemes where some parameters are $\drbar$ while others are OS. The study of
different renormalisation schemes is very important. First it can provide an estimate on the theoretical uncertainty due to the truncation to one-loop of the perturbative prediction and may also point at a bad choice of a
renormalisation scheme. Second, for the NMSSM where a large part of the spectrum has not been seen it is difficult
to {\it predict} which, from the point of view of an OS scheme, are the input parameters that one can use or which
are the masses that will be discovered and measured (precisely) {\it first}. It is therefore wise to be open and
prepare for different possibilities. In particular, our discussion will touch on some important issues regarding the relationship between the underlying parameters at the level of the  Lagrangian and the physical parameters. This will bring up the issue of the reconstruction of the underlying parameters which is very much tied up to the renormalisation scheme and the differences in how we define the counterterms.\\

\noi 
One of our goals has been  to implement our approach in a code for the
automatic generation of one-loop corrected observables and for an easy implementation of the counterterms. We have
relied on \sloops. Therefore this work is also a natural extension of the work performed in
~\cite{Baro:2008bg,Baro:2009gn} for the MSSM. Taking advantage of this automation  we are able to provide and
discuss a series of applications, pertaining to corrections to masses and various decays involving charginos,
neutralinos and sfermions.

\noi The paper is organized as follows. Section~\ref{sec:descripNMMS}
contains a brief description of the NMSSM. 
Our general approach to the renormalisation of the NMSSM and its implementation in {\tt SloopS} as well as how we handle infra-red divergences is explained in Section~\ref{sec:oneloop-general}. 
The renormalisation of the neutralino and chargino sector is detailed in Section~\ref{sec:ren-neut-charg}. We also give a rather extensive presentation of the different choices for the on-shell schemes and the problematic of the choice of the input parameters. The renormalisation of the sfermion sector follows the one of the MSSM. It is briefly reviewed in Section~\ref{sec:ren_sf}. We are then ready to apply the general approach and principles to specific observables. We start in Section~\ref{sec:appli-benchmarks} by defining a set of 5 benchmark points. 
 In Section~\ref{sec:one-loop-results-chi} we first start by giving results for different schemes for the one-loop corrected masses of the neutralinos before presenting results for the one loop corrected two-body decays of charginos and neutralinos into gauge bosons. This is performed for all 5 benchmark points and for different schemes. We then turn in Section~\ref{sec:one-loop-sfermion}
to the one-loop two-body decays of third generation sfermions into a fermion and chargino or neutralino. Section~\ref{sec:conclusions} contains our conclusions.

\section{Description of the NMSSM}
\label{sec:descripNMMS}
The NMSSM contains all the superfields of the MSSM as well as one additional gauge singlet superfield $\hat{S}$.
Thus the Higgs sector consists of  two SU(2) Higgs doublets superfields $\hat{H}_d$,$\hat{H}_u$ and the singlet superfield,
\begin{equation}
\hat{H}_u=\begin{pmatrix}
\hat{H}_u^+\\
\hat{H}_u^0
\end{pmatrix},\quad \hat{H}_d=\begin{pmatrix}
\hat{H}_d^0\\
\hat{H}_d^-
\end{pmatrix},\quad \hat{S}.
\end{equation}

\noi The interaction Lagrangian can be decomposed in terms derived from the superpotential and from the soft SUSY
breaking Lagrangian.
In the $\mathbb{Z}_3$-invariant NMSSM that we consider here, the superpotential can be split into two
parts~\cite{Ellwanger:2009dp}. The first one depends only on the Higgs superfields $\hat{H}_d$,$\hat{H}_u,\hat{S}$
via two dimensionless couplings $\lambda$ and $\kappa$,
\begin{equation}
W_{NMSSM}=-\lambda\hat{S}\hat{H}_d\cdot \hat{H}_u+\frac{1}{3}\kappa\hat{S}^3,
\end{equation}
where $\hat{H}_d \cdot\hat{H}_u = \epsilon_{ab} \hat{H}_d^a \hat{H}_u^b$ and $\epsilon_{ab}$ is the two dimensional
Levi-Civita symbol with $\epsilon_{12} = 1$.
The second part corresponds to the Yukawa couplings between Higgs and quarks or leptons superfields, 
\begin{equation}
W_{Yukawa}=-y_u\hat{H}_u\cdot \hat{Q} \hat{U}^c_R-y_d\hat{H}_d\cdot \hat{Q} \hat{D}^c_R-y_e\hat{H}_d\cdot
\hat{L}\hat{E}^c_R,
\end{equation}
where 
\begin{equation}
\hat{Q}_i=\begin{pmatrix}
\hat{U}_{iL}\\
\hat{D}_{iL}
\end{pmatrix},\quad \hat{L}_i=\begin{pmatrix}
\hat{\nu}_{iL}\\
\hat{E}_{iL}
\end{pmatrix},\quad \hat{U}_{iR},\quad \hat{D}_{iR},\quad \hat{E}_{iR},
\end{equation}
are respectively the superfields associated with the left-handed (LH)  quark doublets, LH lepton doublets, right-handed (RH)
 quark and  lepton singlets. 
The index i=1..3 indicates the generation. In what follows, this index will be omitted and a sum over the three
generations will be implicit. No generation mixing is assumed in our study. 
These supersymmetric scalar partners  will be denoted as 
$\tilde{Q}=\begin{pmatrix}\tilde{u}_L\\\tilde{d}_L\end{pmatrix}$ and
$\tilde{L}=\begin{pmatrix}\tilde{\nu}_L\\\tilde{e}_L\end{pmatrix}$ for the LH states and $\tilde{u}_R,\tilde{d}_R$ and $\tilde{e}_R$ for the partners of the RH states. In an abuse of language we will also refer to these partners as LH and RH.
\noi Let us keep in mind, at this point already, that parameters from the superpotential will  find their way into the Lagrangian of the particle and the superparticles. For example, the same $\lambda, \kappa$ enter both the Higgs
sector {\it and} the neutralino (higgsino) sector, thus offering ways to extract these parameters from different
sectors.
The soft SUSY breaking Lagrangian reads,
\begin{align}
-\mathcal{L}_{soft}&=m_{H_u}^2|H_u|^2+m_{H_d}^2|H_d|^2+m_S^2|S|^2 \notag\\
&  \quad \quad \quad + (\lambda A_{\lambda}H_u\cdot H_dS+\frac{1}{3}\kappa A_{\kappa}S^3+h.c)\notag\\
&+
m_{\tilde{Q}}^2|\tilde{Q}|^2+m^2_{\tilde{u}}|\tilde{u}^2_R|+m^2_{\tilde{d}}|\tilde{d}^2_R|+m^2_{\tilde{L}}|\tilde{L}^2|+m^2_{\tilde{e}}|\tilde{e}^2_R|\notag\\
& \quad \quad \quad +(y_uA_u\tilde{Q}\cdot H_u\tilde{u}^c_R-y_dA_d\tilde{Q}\cdot
H_d\tilde{d}^c_R-y_eA_e\tilde{L}\cdot H_d\tilde{e}^c_R)\notag\\
&-\frac{1}{2}\left(M_1\tilde{B}\tilde{B}+M_2\tilde{W}_i\tilde{W}_i+M_3\tilde{G}^a\tilde{G}^a\right),
\label{softLagr}
\end{align}
\begin{itemize}
\item The first two lines belong to the Higgs sector with the first line representing the soft mass terms for the
Higgs bosons
while the second line, not present in the MSSM, represents the NMSSM trilinear Higgs couplings $A_\kappa,A_\lambda$.
\item The third and fourth lines belong to the sfermion sector with a structure and a content exactly the same as in the MSSM with first the soft sfermion masses ($m_{\tilde{Q}/\tilde{L}}$ for the doublet squark/slepton, and $m_{\tilde{u},\tilde{d},\tilde{e}}$ for the RH singlets) followed by the MSSM-like tri-linear $A$-terms for squarks and sleptons $A_u,A_d$ and $A_e$. We have only written the terms for one generic generation since we are not considering inter-generation mixing.
\item The last line contains the soft mass terms for, respectively, the $U(1)$, $SU(2)$ and $SU(3)$ gauginos, also
called bino, winos and gluinos.
\end{itemize}
\noi We consider the NMSSM with CP conservation so that all parameters are taken to be real. \\

\noi The neutral components of the Higgs doublets, $H_u$ and $H_d$, contain both a CP even and a CP odd part. After
expanding around their vacuum expectation values, their scalar
neutral component reads
\begin{equation}
H_d^0=v_d+\frac{1}{\sqrt{2}}\left( h_d^0+i a_d^0\right) \;, \;\; H_u^0=v_u+\frac{1}{\sqrt{2}}\left( h_u^0+i
a_u^0\right) \;, \;\; S^0=s+\frac{1}{\sqrt{2}}\left( h_s^0+i a_s^0\right)
\end{equation}
The vacuum expectation values, $v_u,v_d,s$ are chosen to be real and positive. As in the MSSM we define
$\tan\beta\equiv t_\beta=v_u/v_d$ and $v^2=v_u^2+v_d^2 $ such that the $W$ mass comes out to be
$M_W^2=g^2v^2/2$. \\

\noi The so-called higgsino mass parameter in the MSSM is now a derived parameter. $\mu$ is generated dynamically
from the vev of the singlet field,
\begin{equation}
\mu=\lambda s .
\end{equation}
It is convenient to keep $\mu$ as an independent parameter, comparison with the  MSSM will then be easier. With
$\mu$, we take $\lambda$ and $\kappa$ as independent parameter while $s$ is kept as a shorthand notation for
$\mu/\lambda$ in the same way as we use $c_W$ as a short-hand notation for $M_W/M_Z$.

\noi The particle content of the NMSSM has extra particles in the neutralino and Higgs sector than what
constitutes the MSSM. The physical scalar fields consist of 3 neutral CP-even Higgs bosons, $h_1^0,h_2^0,h_3^0$, 2 CP-odd Higgs bosons,
$A_1^0,A_2^0$ and a charged Higgs, $H^\pm$.
The fermionic component of $\hat{S}$ is a neutralino called singlino. It mixes with the two higgsinos.  With the two gauginos 
(U(1) and SU(2))  the NMSSM has five neutralinos.\\

\noi To summarise, the parameters that will be relevant for the present paper which covers the neutralino, chargino and
sfermion sector and which need to be renormalised (apart from the SM parameters) are
\begin{equation}
\label{para_count_1}
\underbrace{t_\beta, \lambda,\kappa,\mu}_{{\rm in \; Higgs \; also}}, M_1,M_2 \;\;\; ; \;\;\; m_{\tilde{Q}},
m_{\tilde{u}_R}, m_{\tilde{d}_R}, A_u,A_d \;\; ; \;\; m_{\tilde{L}},m_{\tilde{e}_R}, A_e,
\end{equation}

\noi The first six of these  parameters enter the chargino/neutralino sector.  
$t_\beta, \lambda,\kappa,\mu$ also enter the Higgs sector. In fact $t_\beta$ and $\mu$ are also present in the sfermion sector. The second group corresponds to the squark sector while the last group corresponds to the
sleptons. \\

\noi Other parameters not listed in Eq.~\ref{para_count_1} such as $A_\kappa$ and $A_\lambda$ enter only the Higgs
sector. They will be studied  in a separate publication detailing the treatment of the Higgs sector.
Because of the supersymmetric nature of the model, in particular the origin of the $\mu$ parameter, the
neutralino/chargino sector and the Higgs share parameters in common as was presented in Eq.~\ref{para_count_1}.
Since it may be advantageous to use inputs from the Higgs sector to extract one or all of the parameters $t_\beta,
\lambda,\kappa,\mu$ in Eq.~\ref{para_count_1}, their extraction and definition will be influenced by how all the
parameters of the Higgs sector are extracted. Let us therefore list the 9 parameters of the Higgs sector:
\begin{equation}
\label{para_count_higgs}
\underbrace{t_\beta, \lambda,\kappa,\mu}_{{\rm in} \; \tilde{\chi} \; {\rm sector \; also}}, A_{\lambda}, A_\kappa,
m_{H_d},m_{H_u},m_{S}.
\end{equation}
\\
\noi Finally since we concentrate on electroweak corrections and do not consider gluino production or decay, the
renormalisation of $M_3$ is not needed.

\section{Full one-loop corrections: general approach}
\label{sec:oneloop-general}

\subsection{Renormalisation: our general approach}
\label{subsec:renorm}
The renormalisation procedure follows the same approach as the one adopted in {\tt SloopS} for the SM and the MSSM.
Namely we aim primarily at an on-shell renormalisation of all parameters~\cite{Baro:2008bg,Baro:2009gn}.
Other realisations of on-shell renormalisation schemes for the chargino/neutralino sector have also been performed
both in the MSSM~\cite{Chatterjee:2011wc,Liebler:2010bi}, the complex MSSM~\cite{Bharucha:2012nx} and the
NMSSM~\cite{Liebler:2010bi}. \\

\noi OS schemes mean that one uses as inputs physical observables which are therefore defined when particles taking
part in these observables are physical and on their mass shell. Technically, the easiest and most obvious set of
this type of observables are the masses of the particles themselves. In this case one only exploits the pole
structure of two-point self-energy functions and require that the residue at the pole be unity. One difficulty
occurs when we have mixing between particles sharing the same quantum numbers and therefore transitions from one to
the other are possible. This will occur for Higgses, charginos, neutralinos and sfermions. The OS
conditions mean also that when these physical particles are {\it on their mass shell} these (non-diagonal)
transitions vanish. From another technical point of view this means that in the calculation of scattering amplitudes and decays we should not worry about corrections on the external legs, the wave functions will be automatically
normalized. Recall that at tree-level one starts with the underlying parameters of a  Lagrangian in
terms of current/gauge fields where mixing between these fields is present. We then move to the physical basis where the physical fields are defined. This is achieved by some diagonalising matrices. At one-loop each underlying
parameter is shifted by the addition of a counterterm. There is then a minimum set of conditions to restrict the
form and the value of the counterterm. This shifting of parameters will at one-loop mix some particles. To perform a full definition of a physical particle at one-loop, in our approach, we introduce a matrix of wave functions with
the conditions that when these transitions (containing one-loop plus counterterms) are evaluated OS, all transitions vanish. It is important to stress that it is unnecessary to introduce shifts in the diagonalising matrix that was
used at tree-level. \\

\noi Related to mixing also is the fact that one physical parameter, for example the mass of one neutralino in the
NMSSM, depends on a large number of independent underlying parameters contained, in this case, in the $5\times5$
mixing matrix. For instance, besides the SM parameters, 6 parameters (the first set in Eq.~\ref{para_count_1}),
contribute to  the neutralino mass matrix. In this particular case one needs to solve a system of 6
coupled equations.
This is the reason why the reconstruction of the parameters, or in other words the necessary counterterms, requires
finding the solution to a (large) system of coupled equations. Finding the solutions can be extremely difficult and sometimes impossible from a partial or even total knowledge of the physical parameters. For example, the chargino masses can furnish $M_2, \mu$ but with a $M_2 \leftrightarrow \mu$ degeneracy. If the system of coupled
equations can be split into different independent
subsystems of equations, the extraction of the parameters will be much easier and their evaluations less subject to
uncertainties in the sense of being less sensitive to small variations in the input parameters. Therefore, by
combining different sectors one can work with smaller, independent blocks which are easier or more efficiently
solved. For example, take the set in Eq.~\ref{para_count_1}, $t_\beta$
originates from the Higgs sector and finds it way in all sectors of the NMSSM. As we will illustrate, it is much
easier to get the counterterm for $t_\beta$ from the Higgs sector for which we could revert to a $\drbar$ scheme. In this case this involves a one-to-one mapping between the required counterterm for $t_\beta$ and some simple
evaluation of 2-point functions involving the Higgs. Reverting to the Higgs sector for this particular parameter is
therefore technically much easier than trying to extract all the 6 parameters in the first set of
Eq.~\ref{para_count_1} solely from the neutralino/chargino sector. Moreover by extracting $t_\beta$ from the Higgs
sector we can choose a scheme where one further decomposes the remaining system of the $5 \times 5$ coupled equations into 2 blocks: 2 equations from the chargino sector that will then furnish $\mu, M_2$ and the rest can be determined from the neutralino sector. Another advantage is that we have a much better handle on the extraction of $t_\beta$, indeed
as we stressed and as we will see explicitly the effect of $t_\beta$ on the neutralino/chargino is quite small. In a nutshell, a physical mass $M_\chi$ of a neutralino/chargino is essentially given by a soft mass $M$ with a small correction $\epsilon_m$ which is proportional to $\tb$, such that $M_\chi=M+ \tb \epsilon_m$, then $\tb \propto 1/\epsilon_m$.  
Although we will propose to use the Higgs sector for a definition of $t_\beta$, we will in this first paper be very
brief about the renormalisation of the Higgs, the full renormalisation of the Higgs sector will be detailed in a
forthcoming publication~\cite{renormalization_higgs}. In order to facilitate the comparison with other computations, we will also use a $\drbar$
scheme in which the six parameters of the neutralino/chargino sector are taken as $\drbar$ while on-shell conditions are used for the SM parameters.\\

\noi Leaving aside the issue of $t_\beta$ (where it is defined from), the chargino/neutralino sector through the
masses of the 7 particles it contains could furnish enough input to constrain the set of  6 parameters. There are various choices for the minimal set of inputs. We will propose a few. 
The most appropriate choice of input may depend on the observable considered. For example 
imagine a scenario where $M_1$ is much larger than all other masses. The scheme   with the 3 lightest neutralinos will be quite insensitive to $M_1$ and its counterterm. As long as we
concentrate on correcting observables that are not sensitive to the bino component, this should be fine but clearly
within this scheme we should not expect to make a good prediction to any observable where the bino component plays a role. Similar issues occur with  the singlino. The mention
of the bino and singlino component, or any other component for that matter, raises the issue of how can one weigh
any of these components from a knowledge of masses only. In general this is not possible.
 This is one of the
shortcomings of the OS approach based solely on masses that we will present here. Schemes where one can use a
particular decay of a neutralino which is sensitive to a particular coupling and hence component, {\it in lieu} of a
mass, are possible but they are technically challenging (use of three-point function) and we will not implement this approach in this first publication. \\

\noi To be complete, let us recall that the fermion and gauge sector of the SM is renormalised on-shell which means
that the gauge boson masses are defined from the pole masses and that the electromagnetic coupling, $\alpha$, is
defined in the Thomson limit. One should keep in mind that the scale of the latter, $q^2=0$, is far smaller than the electroweak scale or the masses of the various supersymmetric particles we are dealing with. A running of $\alpha$,
from $q^2=0$ to $q^2=M_Z^2$ brings in about a
$7\%$ correction. \\ 

\noi If a complete and proper renormalisation procedure has been achieved, all observables should be ultra-violet
finite. We always perform this stringent test and check for the absence of ultraviolet divergences. Such divergences arise in loop integrals and are encoded in the parameter $C_{UV}$ defined in dimensional reduction as
$C_{UV}=2/\epsilon -\gamma_E+\ln(4\pi)$ where $\epsilon=4-d$, $d$ being the
number of dimensions and $\gamma_E$ is the Euler constant\footnote{In
{\tt SloopS}, we apply the  constrained differential
renormalisation scheme which has been shown to be equivalent  to the SUSY
conserving
dimensional reduction scheme  \cite{Hahn:1998yk}.}.  Since physical processes must be finite,  we simply check that 
the numerical results, for one-loop corrections to masses or to decay processes, are independent of $C_{UV}$ by
varying the numerical value of $C_{UV}$ from 0 to $10^7$. We require that the numerical results agree up to five or
seven digits (recall that \sloops $\;$ uses double precision). Such tests have proven extremely useful in testing
the code at each step of its implementation. \\
\noi In schemes where at least one parameter is taken to be $\drbar$, a dependence on the renormalisation scale
$\bar{\mu}$ also appears. For all decay processes we have set this scale to the mass of the decaying particle and in calculating corrected masses this scale is set at the tree-level mass of the particle.

\subsection{Infrared and real corrections}
\label{subsec:infrared}
A second test concerns infrared finiteness. Infrared divergences arise in processes involving charged particles in external legs. The regularization of the divergence from the pure loop contribution is done in {\tt FormCalc} by
adding a fictitious mass to the photon ($\lambda$). After adding the real photon emission, the divergence associated with the soft photon emission will exactly cancel that of the pure loop contribution ($1V$)
\begin{equation}
\sigma_{1 V+soft}(s,k_c)=\sigma_{1V}(s,\lambda)+\sigma_{soft}(s,\lambda,k_c)
\label{sommeSoft}
\end{equation}
where $k_c$ is a cut on the energy of the photon  introduced to separate the soft and hard part when
performing the phase space integral for the real emission,
\begin{equation}
\sigma_{1
soft+hard}(s,\lambda)=\sigma_{soft}(s,\lambda,k_c)+\sigma_{hard}(s,k_c).
\end{equation}
To check the convergence we modify the value of $\lambda$. Note that
the $k_c$ dependence should disappear when calculating the sum of the soft and hard part. This check is not
automatized in {\tt SloopS}, one has to calculate the sum of soft and hard emission
for different values of $k_c$ until a plateau is reached. \\

\noi In our calculations of the decays of squarks we have also considered QCD corrections. In all examples we
have considered in
the present paper the genuine non-abelian structure of QCD is not present. For all these cases we adopt
the same procedure for taming the infrared divergences concerning gluons as the one we apply to infrared photons.
For these applications we give the gluon a mass.

\section{Renormalisation of the Chargino and Neutralino sector}
\label{sec:ren-neut-charg}

\subsection{Implementing our general considerations}
\label{sec:ren-neut-charg-general}
Before entering into the details of the chargino/neutralino sector let us review our set-up for the renormalisation of the fermions as fit for the neutralinos and charginos. We will follow almost {\it
verbatim} the implementation in the MSSM carried out in~\cite{Baro:2008bg,Baro:2009gn}. We reproduce the different steps so the reader can follow exactly how we impose our conditions on the different counterterms.

For a Dirac fermionic field $\psi=\begin{pmatrix}
\psi^L\\
\psi^{R\dagger}
\end{pmatrix}$ with a bare mass $M_0$, the kinetic and mass terms of the Lagrangian can be written at tree level as
:
\begin{align}
\label{Mzero}
\mathcal{L}_0^{Dirac}=i\left(\psi_0^R\sigma^\mu\partial_\mu\psi_0^{R\dagger}+\psi_0^{L\dagger}\bar{\sigma}^\mu\partial_\mu\psi_0^{L}\right)-M_0\left(\psi_0^L\psi_0^R+\psi_0^{L\dagger}\psi_0^{R\dagger}\right)\quad
.
\end{align}

\noi When several fermions mix, the mass term $M_0$ simply becomes a matrix.
$M_0$ can involve a large number of
underlying parameters. The mass eigenstates are obtained after diagonalizing the mass matrix with two unitary
matrices $D^R$ and $D^L$,
\beqn
\label{Dmatrices}
\chi_0^R=D^R\psi_0^R,\quad\chi_0^L=D^L\psi_0^L\quad ,
\eeqn
such that 
\beqn
\label{DiagD}
\widetilde{M}=D^RMD^{LT}=\widetilde{M}^T={\rm diag}(m_{\chi_1}, m_{\chi_2}, ...).
\eeqn
We now shift $M_0$ by shifting the parameters of its elements and proceed to shift fields through wave function
normalization, 
\begin{align}
M_0=&M+\delta M\label{shift1}\\
\chi_{i_0}^{R,L}=&\left(\delta_{ij}+\frac{1}{2}\delta Z_{ij}^{R,L}\right)\chi_j^{R,L}\label{DeltaM0}
\end{align}
\noi $M$ and $\chi_i$ are the renormalised matrix and fields respectively and $\chi_i^{R,L}= P_{R,L} \chi_i$ where
$P_{R,L}=(1\pm\gamma_5)/2$ are projection operators.
For a Majorana fermion, as will be the case for the neutralinos, $\psi_0^L=\psi_0^R=\psi$, only one counterterm
matrix is required, likewise one unitary matrix is needed for the diagonalisation of the mass matrix. \\
\noi 
Following~\cite{Baro:2009gn} the renormalised two-point function describing the $ i \to j$ transition can be written in a compact notation,
\begin{align}
\hat{\Sigma}_{ij}(q)=\Sigma_{ij}(q)-P_L\delta m_{ij}-P_R\delta
m_{ji}^*&+\frac{1}{2}(\cancel{q}-m_{\chi_i})\left[\delta Z_{ij}^LP_L+\delta Z_{ij}^{R*}\right]\notag\\
&+\frac{1}{2}\left[\delta Z_{ji}^{L*}P_R+\delta Z_{ji}^RP_L\right](\cancel{q}-m_{{\chi_j}})
\label{self}
\end{align}
including the one-loop self-energy $\Sigma_{ij}(q)$ and the counter-terms $\delta m_{ij}$ that represent the
correction to the element $\widetilde{M}_{ij}$, {\it i.e.} $\delta m_{ij}= D^R \delta M D^{LT}$. We stress again
that we are using the same diagonalising matrices $D^{R,L}$ as those used at tree level. This formula makes it clear
that the mass and wave-function counterterms can be obtained separately from on-shell (OS) conditions.

\noi Using one of the masses $m_{\chi_{i}}$ one can
impose one of the OS conditions on the physical pole mass
\begin{eqnarray}
\label{os_mass}
\widetilde{Re}\hat{\Sigma}_{\tilde{\chi}_i
\tilde{\chi}_i}(q)u_{\chi_{i}}(q)=0 \; {\rm for}\;
q^2=m_{\chi_{i}}^2. \label{fieldselfii}
\end{eqnarray}
$\widetilde{Re}$ means that the imaginary
dispersive part of the loop function is discarded so as to
maintain hermiticity at one-loop. $m_{\chi_{i}}$ is the tree-level mass. Using a mass $m_{\chi_{i}}$ as an input
means that the tree-level mass that is used  in Eq.~\ref{os_mass} receives no correction at one-loop. This 
gives a direct constraint on the $\delta m_{ii}$ element which will be used as one condition  to solve for the
system of equations that define the full set of counterterms. When this full set of counterterms is solved equation
Eq.~\ref{os_mass} is used to calculate the pole mass for the particles that were not used as input, see
\cite{Baro:2009gn} for the algebraic details. Considering the number of coupled equations, finiteness of the
mass(es) derived at one-loop is a highly non trivial test and shows the robustness of our code.
We always perform this finiteness test.

\noi Wave-function renormalisation constants are derived by requiring that 
\begin{itemize}
\item[{\bf {\it i)}}]
the diagonal renormalised 2-point self-energies for $i \ra i$ transitions have residue of $1$ at the pole mass. This pole mass
may get a one-loop correction. For our treatment at one-loop it is
sufficient to impose the residue condition by taking the
tree-level mass. This translates into 
\beqn
& & \displaystyle{\lim_{q^2 \ra m_{{\chi}_{i}}^2}}
\frac{\slashq+m_{{\chi}_{i}}}{q^2-m_{{\chi}_{i}}^2}
\widetilde{Re}\hat{\Sigma}_{{\chi}_i
{\chi}_i}(q)u_{\chi_{i}}(q)= u_{\chi_{i}}(q) \; {\rm and} \;
\nonumber \\
& & \displaystyle{\lim_{q^2 \ra m_{{\chi}_{i}}^2}}
\bar{u}_{\chi_{i}}(q) \widetilde{Re}\hat{\Sigma}_{{\chi}_i
{\chi}_i}(q)\frac{\slashq+m_{{\chi}_{i}}}{q^2-m_{{\chi}_{i}}^2}=
\bar{u}_{\chi_{i}}(q) \label{fieldii}
\eeqn
\item[{\bf {\it ii)}}]To avoid any $ i \ra j, i \neq j$,  transition we impose 
\begin{eqnarray}
\widetilde{Re}\hat{\Sigma}_{{\chi}_i
{\chi}_j}(q)u_{\chi_{j}}(q)=0 \; {\rm for}\;
q^2=m_{\chi_{j}}^2\, ,\, (i \neq j )\; .\label{fieldij}
\end{eqnarray}
\end{itemize}

\subsection{Specialising to the case of the charginos and neutralinos}
\label{sec:ren-neut-charg-detail}
The new fermions in the electroweak sector of the NMSSM are the two charginos, combination of charged winos and
higgsinos as in the MSSM, and the five neutralinos, combination of bino, wino, neutral higgsinos and the singlino.
In the basis
\begin{align}
\psi_c^R=\begin{pmatrix}
-i\tilde{W}^-\\
\tilde{H}^-_d
\end{pmatrix},\quad
\psi_c^L=\begin{pmatrix}
-i\tilde{W}^+\\
\tilde{H}^+_u
\end{pmatrix}
\end{align}
the mass matrix for the charginos reads, 
\begin{align}
X=\begin{pmatrix}
M_2&\sqrt{2}M_Ws_\beta\\
\sqrt{2}M_Wc_\beta&\mu
\end{pmatrix},
\label{charmass}
\end{align}
while for the neutralinos in the basis 
\begin{equation}
\psi_n^{RT}=\psi_n^{LT}=\psi^{0T}= \left(
-i\tilde{B}^0,
-i\tilde{W}_3^0,
\tilde{H}_d^0,
\tilde{H}_u^0,
\tilde{S}^0 \right)
\end{equation}
the mass matrix reads
\begin{align}
Y=\begin{pmatrix}
M_1 & 0 & -M_Zs_Wc_{\beta} & M_Z s_W s_{\beta} & 0 \\
0 & M_2 & M_Zc_Wc_{\beta} & -M_Zc_Ws_{\beta} & 0 \\
-M_Zs_Wc_{\beta} & M_Zc_Wc_{\beta} & 0 & -\mu & -\lambda v s_\beta \\
M_Z s_W s_{\beta} & -M_Zc_Ws_{\beta} & -\mu & 0 & -\lambda v c_\beta \\
0 & 0 & -\lambda v s_\beta & -\lambda v c_\beta & 2\kappa s
\end{pmatrix},
\label{neumass}
\end{align} 
The charginos and neutralinos eigenstates are obtained with the help of two
unitary matrices $U$ and $V$ for charginos
and one unitary matrix $N$ for neutralinos ($U,V,N$ are particular
manifestations of the matrices $D^{L,R}$ introduced in Eq.~\ref{Dmatrices}:
\begin{align}
\chi^R=U\psi^R_c,\quad \chi^L=V\psi^L_c,\quad \chi^0=N\psi^0\quad  
\end{align}
leading to  the mass eigenstates
\begin{align}
\tilde{X}=U^*XV^{\dagger}={\rm diag}(m_{\tilde{\chi}_1^+},m_{\tilde{\chi}_2^+}), \quad
\tilde{Y}=N^*YN^{\dagger}={\rm
diag}(m_{\tilde{\chi}_1^0},m_{\tilde{\chi}_2^0},m_{\tilde{\chi}_3^0},m_{\tilde{\chi}_4^0},
m_{\tilde{\chi}_5^0}).
\label{diag}
\end{align}
Following our program we proceed to shift the underlying parameters. This results in introducing counterterms to the mass matrices
\begin{align}
\delta X=\begin{pmatrix}
\delta M_2 & \delta X_{12} \\
\delta X_{21} & \delta\mu
\end{pmatrix},\quad
\delta Y = \begin{pmatrix}
\delta M_1 & 0 & \delta Y_{13} & \delta Y_{14} & 0 \\
0 & \delta M_2 & \delta Y_{23} & \delta Y_{24} & 0 \\
\delta Y_{13} & \delta Y_{23} & 0 & -\delta \mu & \delta Y_{35}\\
\delta Y_{14} & \delta Y_{24} & -\delta \mu & 0 & \delta Y_{45}\\
0 & 0 & \delta Y_{35} & \delta Y_{45} & \delta Y_{55}
\end{pmatrix},
\end{align}
with, in the chargino case,  
\begin{align}
\left\{
\begin{array}{r c l}
\delta X_{12}=\sqrt{2}s_\beta\delta M_W+\sqrt{2}M_W s_\beta c_\beta^2\frac{\delta t_\beta}{t_\beta},\\
\delta X_{21}=\sqrt{2}c_\beta\delta M_W-\sqrt{2}M_W s_\beta^2 c_\beta\frac{\delta t_\beta}{t_\beta},
\end{array}
\right.
\end{align}
and for the neutralino counterterms
\begin{align}
\label{deltaY}
\left\{
\begin{array}{r c l}
\delta Y_{13}&=&-M_Zs_Wc_{\beta}\left[\frac{1}{2}\frac{\delta M_Z^2}{M_Z^2}+\frac{1}{2}\frac{\delta
s_W^2}{s_W^2}\right]+M_Zs_W\frac{t_{\beta}^2}{(1+t_{\beta}^2)^{3/2}}\frac{\delta t_{\beta}}{t_{\beta}},\\
\delta Y_{14}&=&+M_Zs_Ws_{\beta}\left[\frac{1}{2}\frac{\delta M_Z^2}{M_Z^2}+\frac{1}{2}\frac{\delta
s_W^2}{s_W^2}\right]+M_Zs_W\frac{t_{\beta}}{(1+t_{\beta}^2)^{3/2}}\frac{\delta t_{\beta}}{t_{\beta}},\\
\delta Y_{23}&=&+M_Zc_Wc_{\beta}\left[\frac{1}{2}\frac{\delta M_Z^2}{M_Z^2}+\frac{1}{2}\frac{\delta
c_W^2}{c_W^2}\right]-M_Zc_W\frac{t_{\beta}^2}{(1+t_{\beta}^2)^{3/2}}\frac{\delta t_{\beta}}{t_{\beta}},\\
\delta Y_{24}&=&-M_Zc_Ws_{\beta}\left[\frac{1}{2}\frac{\delta M_Z^2}{M_Z^2}+\frac{1}{2}\frac{\delta
c_W^2}{c_W^2}\right]-M_Zc_W\frac{t_{\beta}}{(1+t_{\beta}^2)^{3/2}}\frac{\delta t_{\beta}}{t_{\beta}},\\
\delta Y_{35}&=&-vs_{\beta}\delta\lambda-\lambda vs_{\beta}c_{\beta}^2\frac{\delta t_{\beta}}{t_{\beta}}-\lambda
s_{\beta}v\left(\frac{\delta M_W}{M_W}-\frac{\delta e}{e}+\frac{\delta s_W}{s_W}\right),\\
\delta Y_{45}&=&-vc_{\beta}\delta\lambda+\lambda vs_{\beta}^2c_{\beta}\frac{\delta t_{\beta}}{t_{\beta}}-\lambda
c_{\beta}v\left(\frac{\delta M_W}{M_W}-\frac{\delta e}{e}+\frac{\delta s_W}{s_W}\right),\\
\delta Y_{55}&=&2(\kappa \delta s+s\delta\kappa).
\end{array}
\right.
\end{align}
with the constraint $\delta c_W^2=-\delta s_W^2=\delta (M_W^2/M_Z^2)$ and $\delta \mu=\delta (\lambda s)$ ($v$ is
also defined from $\alpha, M_W,M_Z$, a constraint which is implemented explicitly in Eq.~\ref{deltaY} ). \\
As we have shown in the general presentation, the
renormalised self energies lead to corrections, $\delta m_{\chi_i}$, to the tree-level masses. Imposing that some of these corrections vanish will put constraints on $\delta X, \delta Y$ or else will give finite one-loop correction
to the mass. Note again that since after the shifts on the parameters are made we still keep the same diagonalising
matrices, we have for the corrections on the physical masses
\begin{align}
{\rm diag}(\delta {m_{\tilde{\chi}_i^\pm}})=\delta\tilde{X}=U^*\delta XV^{\dagger},\quad {\rm diag}(\delta
m_{\tilde{\chi}_i^0})=\delta\tilde{Y}=N^*\delta YN^\dagger\quad .
\label{massRenorm}
\end{align}

\subsection{Issues in the reconstruction of the counterterms of the chargino and neutralino sector}
\label{subsec:reconstruction-chi}
To fully define the chargino/neutralino sector one needs, besides the SM parameters $\alpha$ and $M_{W,Z}$, to
reconstruct and define the  6 parameters  listed in
Eq.~\ref{para_count_1}
namely $t_\beta, \lambda,\kappa,\mu, M_1,M_2$. This set defines the matrices $X,Y$, see
Eq.~\ref{charmass},\ref{neumass}. Three of these parameters are common to both the neutralino sector and the chargino
sector, these are $t_\beta, \mu, M_2$ while $M_1, \lambda, \kappa$ are present only in the neutralino sector.
Clearly  the sole knowledge of two chargino masses is not sufficient to constrain $\mu,M_2$ and
$t_\beta$. 
However, if $t_\beta$ is provided from some other source then input from the two chargino masses can
reconstruct $M_2,\mu$. In this case three neutralino masses are sufficient to define $M_1, \lambda, \kappa$ for this one needs to solve a system of three equations. \\
In principle, the chargino/neutralino sector by providing 7 physical masses can furnish enough constraint to define
the set of the 6 counterterms. However apart from assuming that one is in the lucky situation that as many as  6 (or 7) masses  in the chargino/neutralino sector, have been measured, a cursory look at
the tree-level mass matrices $X$ Eq.~\ref{charmass} and $Y$ Eq.~\ref{neumass} already reveals the problems encountered
in reconstructing the fundamental parameters of these mass matrices from the masses of the charginos and neutralinos only.
First of all,  we see that in the chargino sector, the $t_\beta$ contribution is  quite small. In the neutralino sector
the situation as concerns this parameter is not much better since either its contribution vanishes in the gaugeless
limit ($g \to 0$ or $M_W,M_Z\to 0$), as in the chargino case or it is very much tangled up with the parameter $\lambda$. Moreover both $t_\beta, \lambda$ represent mixing effects that may be difficult to extract from masses only. This is different
from the extraction of $M_1$ for example where {\it if an almost bino-like} neutralino mass,
$m_{{\tilde{\chi}}^{0}_{i}}$, is used as input we would have an almost one-to-one mapping $M_1 \sim
m_{{\tilde{\chi}}^{0}_{i}}$. This said one must not forget that the problematic $t_\beta, \lambda$ are also present
in the Higgs sector and in view of the observations we have made it is worth studying whether some input from the
Higgs sector may not be a better way of extracting $t_\beta, \lambda$. However other parameters enter the Higgs sector but not the chargino/neutralino sector,
see Eq.~\ref{para_count_higgs}. Hence  combining the Higgs  and the chargino/neutralino sectors  as many as 11
parameters should be reconstructed and we would therefore need as many inputs.
\\
\noi We would also like to point at an important conceptual issue having to do with the reconstruction of the
underlying
parameters from the sole knowledge of the physical masses, in particular from the chargino and neutralino sector. As is clear from the chargino mass matrix Eq.~\ref{charmass} there is a $M_2 \leftrightarrow \mu$ symmetry. Although the
system can be solved by giving the two physical chargino masses it is impossible to unambiguously assign the value
of $\mu$ or $M_2$ to the correct ``position" in the mass matrix. In other words the higgsino/wino content is not
unambiguously assigned. This would however be important to know when we want to solve for the other remaining
parameters in the neutralino sector. Even without this caveat a similar problem occurs if one wants to unambiguously extract $M_1$ for example. A good reconstruction would require knowing not only the mass but the bino or singlino
{\it content} of that mass. This is a challenging problem even in the (simpler) MSSM,
~\cite{Kneur:1998gy,Chatterjee:2011wc, Bharucha:2012ya}.
We will assume that some knowledge of the content is available from a measurement of some decay or cross section and from comparing  the chargino and neutralino mass spectrum, see \cite{Baro:2009gn} for a discussion on this
issue. \\
\noi Setting aside these issues and remarks, let us return to the problem of defining and reconstructing the
underlying
parameters and counterterms. Since, for the chargino/neutralino system, we need to define and solve for 6
counterterms, we need a trade-off  that supplies 6 inputs or conditions, ${\tt input}_1,\cdots, {\tt input}_6$.
Different choices of the $n=6$ inputs correspond to a renormalisation scheme. We have also discussed that we may have to revert to a larger set that includes the Higgs sector, in this case solving for both the Higgs and
chargino/neutralino we may have to extend the 6 needed inputs to as many as $n=11$, see
Eq.~\ref{para_count_higgs}. \\
\noi Therefore, in all generality,  one needs to invert a system such as
\begin{align}
\begin{pmatrix}
\delta {\tt input}_1 \\
\cdots\\
\cdots \\
\delta {\tt input}_n\\
\end{pmatrix}={\mathcal{P}}_{n, {\rm param.}} \begin{pmatrix}
\delta\mu\\
\delta M_2\\
\delta \kappa\\
\delta M_1\\
\delta \lambda \\
\delta t_\beta \\
\cdots
\end{pmatrix}+{\mathcal{R}}_{n,\rm residual},
\label{Pnn}
\end{align}
${\mathcal{R}}_{n,\rm residual}$  contains other counterterms, such as gauge
couplings,  that are defined separately. Using the physical mass of one of the
neutralinos/charginos as an input, see Eq.~\ref{os_mass}, is a possible choice 
in an OS scheme. Not all inputs need to be OS. In fact it is perfectly legitimate to adopt a fully
$\drbar$ scheme. In this particular case, the counterterms can be simply read off from an external code such as {\tt
NMSSMTools} or any code based on the solution of the Renormalisation Group
Equation (RGE), at one-loop. In
passing let us add that we have checked systematically that the $C_{UV}$ part of our counterterms are the same,
independently of how we extract them and we checked that they are consistent with the values extracted from {\tt
NMSSMTools}.

\noi To make the system Eq.~\ref{Pnn} manageable one should strive
to reduce the rank of the matrix ${\mathcal{P}}_{n, {\rm param.}}$ by breaking it into independent blocks, such that
\beqn
\mathcal{P}_{n, {\rm param.}} = \mathcal{P}_{m, {\rm param.}} \oplus \mathcal{P}_{p, {\rm param.}} \oplus \cdots,
\quad m+p+\cdots=n
\eeqn
We will compare a few schemes and implementations. In what we will call the mixed $\overline {\textrm{DR}}$ on-shell schemes, we work to reconstruct the 6 parameters of the chargino/neutralino sector, therefore $n=6$. $t_\beta$ will be 
extracted from a $\overline {\textrm{DR}}$ condition on $t_\beta$ (from the Higgs sector), $M_2,\mu$ from the
charginos and the rest of the three parameters solely from the neutralinos. In this case we have

\beqn
\mathcal{P}_{6, {\rm param.} }= \mathcal{P}_{1, {\rm param.}} \oplus \mathcal{P}_{2, {\rm param.}} \oplus
\mathcal{P}_{3, {\rm param.}}
\eeqn
As with all resolutions of a system of equations, the inversion of the matrix $\mathcal{P}$ could introduce the
inverse of a small determinant.  We have already encountered such an example with $\tb$ and the division by the small $\epsilon_{m}$ in section~\ref{subsec:renorm}.
 Another case concerns  $M_1$ that can only be reconstructed precisely using the
neutralino that is dominantly bino. This can easily be seen from the first term in eq.~\ref{massRenorm}, $\delta m_{\tilde{\chi}_i^0} =N_{i1}^{*2}
\delta M_1 + ... $ .
If the mass of the dominantly bino neutralino is not chosen as an input parameter, then the extraction of $\delta
M_1$  involves a division by a small number since  $N_{i1}$ is suppressed, hence can induce numerical instabilities.
 This is the reason we have brought up the issue of the {\it content} of
the particle when its mass is used as input. 
 
\noi A second set of schemes, full OS-scheme, is a full $\mathcal{P}_{6, {\rm param.}} $ where all inputs are masses from
the chargino/neutralino sector. We have pointed at some of the shortcomings of this approach, lack of sensitivity to $t_\beta$ and to $\lambda$ to some extent. To achieve a better determination of the parameters in particular the
problematic $t_\beta$, we get help from the Higgs sector but this time all parameters are defined OS. In this case
among the inputs we will take some Higgs masses. This will be done at the expense of having a larger system,
$\mathcal{P}_{8, {\rm param.} }$, the extra two parameters that come into play are $A_{\lambda,\kappa}$

\subsection{Mixed $\overline {\textrm{DR}}$ on-shell schemes}
\label{subsec:mixed-drbar}
This set up is done along the decomposition $\mathcal{P}_{1, {\rm param.}} \oplus \mathcal{P}_{2, {\rm param.}}
\oplus \mathcal{P}_{3, {\rm param.}}$ where $\mathcal{P}_{1, {\rm param.}}$ gets its source in the Higgs sector,
implementing a $\overline {\textrm{DR}}$ condition for $t_\beta$.

\subsubsection{$t_\beta$ from the Higgs sector}
\label{subsunsec:tb-higgs}
The renormalisation of the Higgs sector  is done within the same spirit as the one followed for the neutralino sector by the introduction of wave function renormalisation constants, details will be given in a separate paper.
 The $\overline{\textrm{DR}}$ condition calls for the wave function
renormalisation constants of the Higgs doublets. It is an extension of the DCPR
scheme\cite{Chankowski:1992er,Dabelstein:1994hb} used in the context of the MSSM to the NMSSM\cite{Ender:2011qh},
\begin{align}
\delta t_\beta=\left[\frac{t_\beta}{2}(\delta Z_{H_u}-\delta Z_{H_d})\right]_\infty,
\end{align}
where $\delta Z_{H_u}$ and $\delta Z_{H_d}$ are the wave function renormalisation constants of the $H_u$ and $H_d$
doublets. The infinity symbol indicates that we take the divergent part of the expression. $\delta Z_{H_u}$ and
$\delta Z_{H_d}$ are related to the wave function renormalisation constants $Z_{h_i h_i}$ of the physical CP-even
eigenstates $h_1^0$, $h_2^0$ and $h_3^0$. The latter are obtained from the CP-even neutral elements of $H_u$ and
$H_d$ through the diagonalising matrix $S_h$
\begin{equation}
(h_1^0, h_2^0, h_3^0)=
(h_d^0, h_u^0, h_s^0) S_h^T
\end{equation}
Explicitly,
\begin{equation}
\delta Z_{H_d}=\frac{1}{R}\sum_{i,j,k=1}^3\epsilon_{ijk}S_{h,j3}S_{h,k2}\delta Z_{h_ih_i}, \quad
\delta Z_{H_u}=\frac{1}{R}\sum_{i,j,k=1}^3\epsilon_{ijk}S_{h,j1}S_{h,k3}\delta Z_{h_ih_i},
\end{equation}
with 
\begin{equation}
\delta Z_{h_ih_i}=\Sigma_{h_ih_i}'(m_{h_i}^2), \quad \quad
R=-\sum_{i,j,k=1}^3\epsilon_{ijk}S_{h_i1}^2S_{h,j2}^2S_{h,k3}^2,
\end{equation}
where $\epsilon_{ijk}$ is the fully antisymmetric rank 3 tensor with $\epsilon_{123}=1$ and
$\Sigma_{h_ih_i}'(m_{h_i}^2)$ is the derivative of the self-energy of the Higgs
$h_i$ (with respect to its external momentum), evaluated at its mass
$m_{h_i}$, this condition is such that the residue of the Higgs propagator is unity. The same requirement was
imposed on the charginos and neutralinos. \\

\noi In a $\overline{\textrm{DR}}$ scheme only the divergent part of the countertem is defined \textit{i.e}, any
finite term is set to $0$. Nonetheless, the scheme and the one-loop result is still not fully defined unless one
specifies the renormalisation scale $\bar{\mu}$. The latter is the remnant scale introduced by the regularization
procedure, dimensional reduction. Varying $\bar{\mu}$ can give some estimate on the theoretical uncertainty of the
calculation due to the truncation at one-loop. In the numerical results obtained using a
$\overline{\textrm{DR}}$ scheme, the default value of $\bar \mu$ is fixed to be equal to the mass of the decaying
particle or to the (tree-level) mass of the particle whose one-loop correction is calculated.

\subsubsection{The charginos}
\label{subsubsec:charginos}
Having solved for $t_\beta$, the chargino system, $\mathcal{P}_{2, {\rm param.}}$ provides the simplest set up for
defining $\mu, M_2$ from the masses of both charginos as input. Exactly the same approach and the same expressions are
found for the MSSM
\begin{align}
\delta M_2=&\frac{1}{M_2^2-\mu^2}\left((M_2m_{\tilde{\chi}_1^+}^2-\mu \det X)\frac{\delta
m_{\tilde{\chi}_1^+}}{m_{\tilde{\chi}_1^+}}+(M_2m_{\tilde{\chi}_2^+}^2-\mu\det X)\frac{\delta
m_{\tilde{\chi}_2^+}}{m_{\tilde{\chi}_2^+}}\right.\notag\\
&\left.-M_W^2(M_2+\mu s_{2\beta})\frac{\delta M_W^2}{M_W^2}-\mu M_W^2c_{2\beta}s_{2\beta}\frac{\delta
t_{\beta}}{t_{\beta}}\right)\quad ,\notag\\
\delta\mu=&\frac{1}{\mu^2-M_2^2}\left((\mu m_{\tilde{\chi}_1^+}^2-M_2 \det X)\frac{\delta
m_{\tilde{\chi}_1^+}}{m_{\tilde{\chi}_1^+}}+(\mu m_{\tilde{\chi}_2^+}^2-M_2\det X)\frac{\delta
m_{\tilde{\chi}_2^+}}{m_{\tilde{\chi}_2^+}}\right.\notag\\
&\left.-M_W^2(\mu+M_2 s_{2\beta})\frac{\delta M_W^2}{M_W^2}-M_2M_W^2s_{2\beta}c_{2\beta}\frac{\delta
t_{\beta}}{t_{\beta}}\right)\quad ,
\label{charg_invert}
\end{align}

\noi The explicit solutions shown in Eq.~\ref{charg_invert} gives us the opportunity to go over  the
ambiguity on the {\it true} reconstruction of $M_2,\mu$. In fact Eq.~\ref{charg_invert} corresponds to 4 solutions,
since $M_2,\mu$ are given up to a sign and since we have a $M_2 \leftrightarrow \mu$ ambiguity. This issue was
discussed at some length and some suggestions were given on how to lift the degeneracy~\cite{Baro:2009gn}. By looking
at the values of some decays (or cross sections) involving a chargino, for example, we can
check that only one of the solutions is compatible with the value of the decay rate. This is a limitation on using
only the value of the physical masses as input. Having chosen the correct $\delta \mu, \delta M_2$ we can now pass
them to the neutralino sector
\footnote{Numerical problems may arise in
the limit $\mu=M_2$, see ~\cite{Baro:2009gn} for a more thorough discussion.}.

\subsubsection{Three neutralino masses as input}
\label{3neut_mass_sc}
We are now left with determining $\delta M_1$, $\delta\kappa$ (or $\delta (\kappa s)$) and $\delta \lambda$ using
three
neutralino
masses, this is the $\mathcal{P}_{3, {\rm param.}}$. Out of the five possible neutralino masses, assuming they have
all been
measured, one must pick up 3 masses that give the best reconstruction of the remaining parameters. As we pointed
out,
technically we should avoid having ${\rm Det}(\mathcal{P}_{3, {\rm param.}}) \to 0$. Obviously the best extraction
of $M_1$
would, ideally, need the {\it bino} like neutralino, whereas $\delta\kappa$
 (or $\delta (\kappa s)$) is most directly tied up with the
singlino component. A wino-like neutralino as a third input will not do
since this is essentially sensitive to
$M_2$ with
only  feeble mixing with the $\lambda$ contribution. The third neutralino to use as input is necessarily a
{\it  higgsino} like neutralino, again this is evident since $\lambda$ in the NMSSM is intimately related to $\mu$,
the higgsino parameter. One can also look at the mass matrix (Eq.~\ref{neumass}) to see that $\lambda$
enters only in the singlino - higgsino off-diagonal element. Therefore the subset to choose calls for $ \delta
m_{\tilde{\chi}_{``{\rm singlino}"}^0} ,
\delta m_{\tilde{\chi}_{``{\rm bino}"}^0}$ and $\delta m_{\tilde{\chi}_{``{\rm higgsino}"}^0}$.
We see again that a judicious choice calls for a knowledge of the {\it identity} of the particle apart
from knowing the value of the corresponding mass exactly. \\
 \noindent Having implemented the 
$\mathcal{P}_{1, {\rm param.}} \oplus \mathcal{P}_{2, {\rm param.}} \oplus \mathcal{P}_{3, {\rm param.}}$ approach
this way, one can calculate the one-loop corrections to two neutralinos, the remaining wino-like and the remaining
higgsino-like neutralinos.

\subsection{Full OS-schemes}
\subsubsection{The neutralino/chargino sector}
\label{subsec:fullos-chi}
Since all 6  parameters $t_\beta, \lambda,\kappa,\mu, M_1,M_2$ are necessary to describe the chargino/neutralino
sector which
provides $7$ physical masses one could entertain defining all these parameters
from this sector. We have pointed out at the shortcomings of this extraction
which has to do with the fact that the dependence on $t_\beta$ is very weak
and that the dependence on
$\lambda$ is complicated. From the technical point of view the reconstruction is also involved as it requires
inverting a $6 \times 6$  system, $\mathcal{P}_{6, {\rm param.}}$. The best choice for
$\mathcal{P}_{6, {\rm param.}}$ builds up on the remarks we
have just made in picking up the three most appropriate neutralinos in the previous paragraph. Based on those arguments
the $\mathcal{P}_{6, {\rm param.}}$ OS scheme uses the following set of inputs

\begin{align}
\label{P6}
\begin{pmatrix}
\delta m_{\tilde{\chi}_1^\pm}\\
\delta m_{\tilde{\chi}_2^\pm}\\
\delta m_{\tilde{\chi}_{``{\rm singlino}"}^0}\\
\delta m_{\tilde{\chi}_{``{\rm bino}"}^0}\\
\delta m_{\tilde{\chi}_{``{\rm higgsino}"}^0}\\
\delta m_{\tilde{\chi}_{``{\rm higgsino}"}^0}\\
\end{pmatrix}=\mathcal{P}_{6, {\rm param.}}
\begin{pmatrix}
\delta\mu\\
\delta M_2\\
\delta \kappa\\
\delta M_1\\
\delta \lambda \\
\delta t_\beta
\end{pmatrix}+\mathcal{R}_{6} \quad ,
\end{align}
In the above we have ordered the inputs in correspondence with the countertems they affect most directly, with the
proviso that the
higgsinos do not reconstruct $t_\beta$ and $\lambda$ efficiently. 

\subsubsection{The  neutralino/chargino and Higgs sectors}
\label{subsec:fullos-chi-higgs}
To improve the determination of $\lambda$ and possibly $\tb$ while keeping with a full OS scheme one has to get help from the Higgs
sector. In that sector
the nature of the mixing between the scalar Higgses means that there is not a one-to-one mapping between $t_\beta$
and a
single Higgs mass. $t_\beta$ gets tangled up with a reconstruction of $A_\kappa$ and $A_\lambda$ which are not
needed for the chargino/neutralino sector. Therefore, at least three Higgs masses are needed.
 The most
natural Higgs masses for this set up, directly related to $A_\kappa$ and $A_\lambda$, are the two
pseudoscalar masses $m_{A_1^0},m_{A_2^0}$. To these one can add the charged Higgs, $H^\pm$, or one of the neutral
CP even Higgsses. In any case the addition of two more inputs for a better determination of the whole set of the
chargino/neutralino sector means we are dealing with $\mathcal{P}_{8, {\rm param.}}$. 
One can also  appeal to the Higgs sector for
 a better determination of $\lambda$  trading another
(second) CP even Higgs for a higgsino-like neutralino.
 Summarising these observations, the variants of  the full OS scheme to extract the counterterms for $\tb,\lambda,\kappa,\mu,M_1,M_2,A_\lambda,A_\kappa$ use
the masses 
$$m_{\tilde{\chi}_1^\pm},  m_{\tilde{\chi}_2^\pm}, \;\; m_{\tilde{\chi}_{``{\rm singlino}"}^0},
m_{\tilde{\chi}_{``{\rm bino}"}^0}, 
( m_{\tilde{\chi}_{``{\rm higgsino}"}^0} \textrm{or}\; m_{h_i^0}),  (m_{H^\pm} \textrm{or}\; m_{h_j^0}), m_{A_1^0},
m_{A_2^0}$$
We refrain from giving the complete formulae for this set up since it relies heavily on the details of the
implementation of the renormalisation of the Higgs sector which will be presented elsewhere~\cite{renormalization_higgs}. Although using an OS approach with the help of Higgs masses can constrain the
singlino parameters we should not expect to have a very good determination of $\tb$. Indeed, even in the MSSM limit
we have shown\cite{Baro:2008bg} that if one takes the heavy CP-even Higgs mass, $M_{H^0}$ as input together with the
pseudo-scalar from the doublet, $M_{A^0}$, then when $M_{A^0} \gg M_Z$
\begin{equation}
\frac{\delta \tb}{\tb} \sim \frac{1}{M_{H^0}^2/M_{A^0}^2-1} \Big( -\delta M_{A^0}^2/M_{A^0}^2 +\delta
M_{H^0}^2/M_{H^0}^2 \Big)
\end{equation}
This could lead to a large finite part when $M_{H^0} \sim M_{A^0}$ as occurs in the decoupling limit.

\section{Renormalisation of the sfermionic sector}
\label{sec:ren_sf}
We now deal with the determination of  the last set of the parameters listed in Eq.~\ref{para_count_1} concerning the sfermion sector. Since the implementation of the sfermionic sector in the NMSSM is exactly the same as in the
MSSM, we have followed  the same approach as the one we developed in~\cite{Baro:2009gn}. We therefore refer
to ~\cite{Baro:2009gn} for details and only summarise the set up here.

\subsection{Squarks}
\label{subsec:sq}
For each generation, 5 parameters, $m_{\tilde{Q}}, m_{\tilde{u}_R}, m_{\tilde{d}_R}, A_u,A_d$, need to be defined (or renormalised) in the squark sector. Recall that each quark $q$ ($q=u,d$) has two scalar superpartners, one for each
chirality, $\tilde{q}_L$ and $\tilde{q}_R$. The squark mass matrix encodes the elements one needs to renormalise. In the $(\tilde{q}_L, \tilde{q}_R)$ basis, the mass matrix $\mathcal{M}_{\tilde{q}}^2$ takes the form (see also
Eq.~\ref{softLagr})
\begin{equation}
\label{sq-matrix}
\mathcal{M}_{\tilde{q}}^2=\begin{pmatrix}
 m_{\tilde{Q}}^2+m_q^2+c_{2\beta}(T^3_q-Q_qs_W^2)M_Z^2& m_q (A_q-\mu t_\beta^{-2T_q^3})\\
 m_q (A_q-\mu t_\beta^{-2T_q^3})& m_{\tilde{q}}^2+m_q^2+c_{2\beta}Q_qs_W^2M_Z^2
\end{pmatrix}.
\end{equation}
where $T_q^3$ is the third component of the isospin for $q$ whose mass is $m_q$.
To define the physical eigenstates, we introduce the diagonalising matrix $R$
such that
\begin{equation}
\begin{pmatrix}
\tilde{q}_1\\\tilde{q}_2
\end{pmatrix}=R\begin{pmatrix}\tilde{q}_L\\\tilde{q}_R\end{pmatrix},\quad R=\begin{pmatrix}
c_{\theta_q}&s_{\theta_q}\\
-s_{\theta_q}&c_{\theta_q}
\end{pmatrix},
\label{rotationMatrix}
\end{equation}
The mass eigenstates will be denoted as $\tilde{q}_{1,2}$ with masses 
\begin{equation}
{\rm diag}(m_{\tilde{q}_1}^2,m_{\tilde{q}_2}^2)=R\mathcal{M}_{\tilde{q}}^2R^T.
\end{equation}
We will take $\tilde{q}_1$ to be the lightest eigenstate. 

We then follow exactly the same procedure as in the neutralino/chargino sector. Namely we shift the underlying
parameters in the mass matrix (Eq.~\ref{sq-matrix}) and introduce wave function renormalisation for the fields
\begin{align}
\mathcal{M}_{\tilde{q}}^2&=\mathcal{M}_{\tilde{q}}^2+\delta\mathcal{M}_{\tilde{q}}^2,\\
\tilde{q}_{i}&=(\delta_{ij}+\frac{1}{2}\delta Z_{ij}^{\tilde{q}})\tilde{q}_j.
\label{shifts_sq}
\end{align}
The ensuing renormalised self-energies for the squarks  read
\begin{equation}
\hat{\Sigma}_{\tilde{q}_i\tilde{q}_j}(q^2)=\Sigma_{\tilde{q}_i\tilde{q}_j}(q^2)-\delta
m_{\tilde{q}_{ij}}^2+\frac{1}{2}\delta Z_{ij}^{\tilde{q}}(q^2-m_{\tilde{q}_i}^2)+\frac{1}{2}\delta
Z_{ji}^{\tilde{q}}(q^2-m_{\tilde{q}_j}^2)
\end{equation}
As was the case in the neutralino/chargino sector, the rotation matrices $R$,  Eq.~\ref{rotationMatrix}, are not
renormalised. This means that the counterterms $\delta m_{\tilde{q}_{ij}}^2$ of the physical mass matrix are given
by :
\begin{equation}
\delta m_{\tilde{q}_{ij}}^2=(R\delta\mathcal{M}_{\tilde{q}^2}R^T)_{ij}
\label{deltaMsquark}
\end{equation}
Keeping with our general strategy we forbid mixing between different fields when they are on their mass-shell,
$Re\hat{\Sigma}_{\tilde{q}_i\tilde{q}_j}(m_{\tilde{q}_i}^2)=0$. Furthermore we set the residue of the renormalised
propagators to unity, $Re\hat{\Sigma}_{\tilde{q}_i\tilde{q}_i}'(m_{\tilde{q}_i}^2)=0$.

\noi Because $SU(2)$ symmetry imposes a common mass to two of the 4 squarks
(before mixing), in our scheme we take 3 physical squark masses as input. In
\sloops\,the selected squark masses  are $m_{\tilde{d}_1}$, $m_{\tilde{d}_2}$
and $m_{\tilde{u}_1}$. The definition of the mixings is directly related to physical observables
namely the amplitude describing the decays $\tilde{u}_2\rightarrow\tilde{u}_1 Z^0$ and
$\tilde{d}_2\rightarrow\tilde{d}_1 Z^0$. At tree-level this amplitude is a substitute for the mixing parameter
$\theta_q, q=u,d$,
${\cal {M}}^{{\tilde q}_2 {\tilde q}_1 Z}=i g_Z T^3_q \sin(2 \theta_q) /2$. $\theta_q$ defined this way is then
promoted to the status of a physical observable ($g_Z=e/s_W c_W$ is extracted from the gauge sector). Therefore the
other two input parameters are $\theta_{u,d}$ for which a counterterm can be defined as
\begin{equation}
\delta m_{\tilde{q}_{12}}^2=-
Re\Sigma_{\tilde{q}_1\tilde{q}_2}\left(\frac{m_{\tilde{q}_1}^2+m_{\tilde{q}_2}^2
}{2}\right),
\label{dmq12}
\end{equation}
see~\cite{Baro:2009gn} for details. These inputs and conditions allow to construct the counterterms for the 5
underlying parameters of the squark sector (Eq.~\ref{para_count_1}). Among the many predictions is that the mass of
the squark
$\tilde{u}_2$ receives a one-loop correction. UV finiteness of this correction is another test of our
implementation.

\subsection{Sleptons}
\label{subsec:sl}
The renormalisation of the slepton sector follows the same methodology and can be considered as a simpler case of
the squark system. Indeed the absence of right-handed neutrinos means that for each generation there are only 3
associated particles: 2 charged sleptons and one sneutrino. Mixing occurs only in the charged sector. 3 parameters,
for each family, need to be fixed,
$m_{\tilde{L}}, m_{\tilde{e}_R}, A_e$, see Eq.~\ref{para_count_1}. The physical masses are $\tilde{e}_1$,
$\tilde{e}_2$ and $\tilde{\nu}$. A simple OS scheme is to take the physical masses of these three particles as input parameters, $m_{\tilde{e}_1}$, $m_{\tilde{e}_2}$ and $m_{\tilde{\nu}}$. An alternative scheme is to take the
(two) charged slepton masses as input with the addition of a constraint on the mixing as we have done for the squark sector, namely $\delta m_{\tilde{e}_{12}}^2=-
Re\Sigma_{\tilde{e}_1\tilde{e}_2}\left(\frac{m_{\tilde{e}_1}^2+m_{\tilde{e}_2}^2}{2}\right)$ as could be extracted
form $\tilde{e}_2 \to \tilde{e}_1 Z$, see ~\cite{Baro:2009gn}. We will stick with the first scheme that requires the
three slepton masses. These different implementations for the squarks and sleptons may be useful when comparing the scheme dependence of the results for sfermion decays.

\section{Benchmark points and  definition of the schemes}
\label{sec:appli-benchmarks}
To apply our formalism we obviously need to fully define a model. In particular our OS renormalisation requires
physical input parameters and most importantly, as we saw, the use of a set of physical masses among the full spectrum. 
 Since no particle of the NMSSM has been discovered yet it is difficult, even within a particular
NMSSM scenario, to pick up the minimal set of input masses. Moreover, even after agreeing on a minimal set to carry the
renormalisation, the other parameters of the model are still needed in order to
perform a complete calculation (some particles and their parameters will only
enter indirectly through their loop effects). The
reason we insist on this seemingly obvious point is that had a particular manifestation of the NMSSM been discovered experimentally, we would have had to use the physical observables, such as some of the physical masses, to
reconstruct the underlying parameters of the model. Such an inversion is notoriously complicated even when 
performed at tree-level, see \cite{Kneur:1998gy} for instance and the discussion for the counterterms in Section~\ref{subsec:reconstruction-chi}. The reconstruction would be easier if information on some
decays and cross sections were given ~\cite{Baro:2008bg}.
The best we can do is the following. We generate models by supplying all the needed underlying parameters, such as
$M_1,M_2, \cdots, A_b, \cdots$. These parameters can be considered as parameters at the electroweak scale. Tree-level formulae
are used to calculate the full spectrum. In turn, for  one-loop calculations, masses of a subset of this spectrum are used as {\it {physical}} masses. The other masses  will receive a loop correction. For example, we can take 3
neutralino masses as input and predict the one-loop corrections for the remaining two neutralinos of the NMSSM.
Another  related issue is that these theory generated (physical) masses from a``known" set of underlying
parameters introduce a bias in our analysis in the sense that we know what the composition of the neutralino is. In
particular from the mass alone one cannot distinguish the singlino-like or bino-like neutral state. What we want to
stress here is that despite our OS approach we have some {\it insider's knowledge} due to the way we generate the
points. This is the reason we will talk about a bino-dominated neutralino for example,  an information easily accessed
through the underlying parameters but much harder to assess from the mass spectrum.
For the same model we will consider different schemes. These correspond to different choices of the input masses for example.

\subsection{Choice of the benchmark points}
\label{subsec:choiceP}
\begin{table}[hp]
\begin{center}
\caption{Parameters for the five benchmark points and tree-level masses of the neutralinos, charginos and third
generation sfermions. For all points, $m_{D_{1,2}}=m_{Q_{1,2}}= m_{L_{1,2}}=A_b=A_\tau=1000$ GeV.
Parameters with mass dimension are expressed in GeV.  }
\label{tab:bench}
\begin{tabular}{|c|c|c|c|c|c|}
\hline
Parameter  &Point1 &Point 2 &Point 3&Point4 &Point 5\\
\hline
$t_\beta$&10&4.5&10&7&3.4\\
$\mu$&250&250&120&600&550\\
$M_1$&1000&230&700&140&400\\
$M_2$&150&600&1000&200&150\\
$M_3$&2500&1000&1000&1000&1000\\
$\lambda$&0.1&0.2&0.1&0.03&0.4\\
$\kappa$&0.1&0.05&0.1&0.007&0.1\\
$A_\lambda$&150&1250&150& 1000&1800\\
$A_\kappa$&  0 &0   & 0  &  0 &0  \\
$A_t$&3000&2200&4000&2300&2400\\
$m_{\tilde{Q}_3}$&2000&1500&2000&1600&1500\\
$m_{\tilde{U}_3}$&2000&500&1000&400&500\\
$m_{\tilde{D}_3}$&1000&1000&1000&1000&1000\\
$m_{\tilde{L}_3}$&1000&1000&1005&1000&1001.5\\
$m_{\tilde{R}_3}$&1000&149.5&1000&140&1005\\\hline
$m_{\tilde\chi_1^0}$&  125.7      & 123.4    & 112.8   & 138.1  & 139.4   \\
$m_{\tilde\chi_2^0}$&   257.3    &  200.9   &   123.8  &  193.1 &   276.2 \\
$m_{\tilde\chi_3^0}$&   278.7      &    255.7  &   241.6 &   280.0 &  392.7  \\
$m_{\tilde\chi_4^0}$&   500.8     &   271.7  &   702.8 &  603.8 &   557.3 \\
$m_{\tilde\chi_5^0}$&    1002.2    &   614.8 &   1006.6 &  612.6 &   574.1 \\
$m_{\tilde\chi_1^+}$&    126.8    &   239.6  &   118.0 &  192.9 &   140.1  \\
$m_{\tilde\chi_2^+}$&     285.9   &   614.7  &   1006.6 &  612.8 &  564.0 \\
$m_{\tilde{t}_1}$& 1873.0 & 459.7 & 935.4 & 358.3 & 453.4\\
$m_{\tilde{t}_2}$& 2132.9 & 1531.7&  2045.1 & 1627.5 & 1533.7\\
$\sin \theta_t$& 0.707 &  0.984 & 0.976  &  0.988&  0.983\\
$m_{\tilde{b}_1}$& 1000.3 & 1000.3 & 1000.3 & 1000.2 & 1000.3 \\
$m_{\tilde{b}_2}$& 2000.9 &  1501.1& 2000.9 &  1601.1&  1501.0\\
$\sin \theta_b$&  1 &  1&  1& 1 & 1 \\
$m_{\tilde{\tau}_1}$& 999.7 & 155.2 & 1000.9 & 146.4 & 1000.8 \\
$m_{\tilde{\tau}_2}$& 1002.3 & 1001.0 & 1006.1 & 1001.1 &1006  \\
$m_{\tilde{\nu}_\tau}$&  998.0 &  998.1&  1003&  998.0&  998.3\\
$\sin\theta_\tau$& 0.727& 1& 0.997& 1& 0.153\\
 \hline
\end{tabular}
\end{center}
\end{table}
We choose five benchmark points in order to cover various hierarchies in the neutralino sector. In particular the
points we selected are
classified according to the nature of the Lightest Supersymmetric Particle (LSP) neutralino. The crucial parameter
that defines the properties of the singlino component of the neutralino is $\lambda$, it ranges from a small value $0.03$ 
(Point 4) to moderate values of the order of the weak gauge coupling $0.1-0.4$. The other parameter that defines the singlino and controls its mass, $m_{\tilde{S}}$, is $\kappa$. 
It is chosen to cover the range $2\kappa/\lambda=0.5-2$. $2 \kappa/\lambda$ is roughly the ratio between the singlino and higgsinos masses.
Sfermions masses of the first two generations as well as the right-handed sbottom are  $\approx 1$ TeV for all the 5 benchmark points. While the mixing for the sbottom is always tiny leading to $\tilde{b}_1=\tilde{b}_R$, we take
large mixings for the stops. Three benchmark points have the lightest stop with mass around 0.5 TeV. Two scenarios
have rather light $\tilde{\tau}_1$ of about $150$ GeV. The LSP's are in the narrow range $110-140$ GeV. 
The values for the underlying parameters for each of the benchmarks are summarised in Table~\ref{tab:bench}. The
parameters of the NMSSM that do not appear in Table~\ref{tab:bench} take a common value for all points,
$A_b=A_\tau= m_{L_{1,2}}=m_{\tilde{D}_{1,2}}=m_{\tilde{Q}_{1,2}}=1000$ GeV while the SM parameters are fixed to
$\alpha=1/137.06, M_Z=91.188 ~\textrm{GeV}, s_W=0.481, \alpha_s(M_Z)=0.118$. 
To summarise
\begin{itemize}
\item Point 1 features a wino-like neutralino LSP. This point exhibits the largest wino-higgsino mixing among all
five scenarios, note that $\mu-M_2 \sim M_Z$.
\item Point 2 has a  singlino dominated LSP. It is also the scenario where the singlino mixings to the other components, while still quite small, are the largest of all 5 scenarios. It also features the largest bino-higgsino
mixing, observe that here $\mu-M_1 \sim M_W/4$, a property that will enhance the bino-higgsino mixing.
\item Point 3 features a   higssino-like LSP.
\item Point 4 has a bino-like  LSP. The singlino is practically decoupled with a very small value of
$\lambda$.
\item Point 5 also has wino-like LSP  but  differs from Point 1 in that the higgsinos are the
heaviest neutralinos. The lightest $\tau$ is mxed though dominantly left-handed. 
\end{itemize}

\noi We have also ensured that these benchmarks were phenomenologically viable, that is they possess a Higgs boson in the 122-128 GeV mass range (after including all loop corrections provided by \texttt{NMSSMTools}) and they satisfy 
theoretical and experimental constraints implemented in \texttt{NMSSMTools}. 
The SM-like Higgs is always the lightest CP-even scalar and all these points
also feature a light pseudoscalar particle
in the range 4-60 GeV. This particle is however not directly relevant for the
numerical examples that follow.
We have also checked that all points satisfy at least the upper bound on the relic density extracted from Planck,
$\Omega h^2 < 0.131$ after taking into account a 10\% theoretical uncertainty~\cite{Ade:2013zuv}.
Points 1,3,5 have a value for the relic density below this range, as typical of wino and higgsino DM below the TeV
scale while Points 2 and 4 fulfill the Planck condition.
To achieve this, we required substantial coannihilation with sfermions, by adjusting $m_{\tilde{R}_3}$ the soft mass term for right-handed sleptons since, typically, scenarios with bino or singlino LSP lead to too much dark matter.

\noi The components (bino, wino, higgsino, singlino) of the neutralinos are shown in Table~\ref{tab:compo}. The neutralinos are labeled from lightest $\neuto$ to heaviest $\neutfi$.  
Since for most points there is not a large
mixing between the components, in order to capture the main properties of the benchmark point at a glance, we will refer to the benchmark in terms of its largest components as $(\neuto,\neutt,\neuth,\neutf,\neutfi) \sim
(\tilde{W}^0_3,\tilde{H}^0,\tilde{H}^0,\tilde{S}^0,\tilde{B}^0)$. 
Note that in the gaugeless limit, $g \to 0
(M_Z,M_W \to 0)$, mixing occurs only between a singlino and a higgsino, the strength of the latter being measured by $\lambda$. In the MSSM limit ($\lambda$ is small) and provided $|M_{1,2}-\mu| > M_Z$, the mixing between the wino and
the higgsino is of the order $M_W/{\rm Max}(\mu,M_2)$ and the mixing between the bino and the higgsino is of order
$M_Z s_W/{\rm Max}(\mu,M_1)$. In the same limit, the mixing between the bino and wino is vanishingly small, this mixing will first
transit {\it via} a higgsino. The $\tb$ dependence is weak, for example in the chargino case the dependence
is hidden in the small mixing factor $M_W (\mu+M_2/\tb)/{\rm Max}(\mu^2,M_2^2)$ and/or $M_W (M_2+\mu/\tb)/{\rm
Max}(\mu^2,M_2^2)$ . 
These general observations explain the values of the mixing in Table~\ref{tab:compo}.
In particular, the largest mixings occur for Point 1 between the wino and the higgsinos and for Point 2 between the
bino and the higgsinos. Point 2  is also the point where the singlino component may be relevant for some of the states
(apart from the LSP singlino, of course). Point 3 and 4  are the ones where the  all neutralinos are the ``purest".
\begin{table}[htb]
\begin{center}\caption{Components of neutralino mass eigenstates for the 5 benchmark points. The dominant component
is highlighted. }
\label{tab:compo}
\begin{tabular}{ccccccc}
\hline
&&Point 1 &Point 2&Point 3&Point 4&Point 5\\
\hline
$\tilde{\chi}_1^0$&$\tilde{B}^0$& - &0.63\%&-&{\bf 98.8\%}&-\\
                           &$\tilde{W}^0$& {\bf  78.6\%}& -&-&- &{\bf 96.2\%}\\
			 &$\tilde{h}^0$& 21.4\% &3.88\%&{\bf 98.4\%}&0.85\%&3.31\%\\
			 &$\tilde{S}^0$&  - &{\bf 95.4\%}&0.77\%&-&-\\\hline
$\tilde{\chi}_2^0$&$\tilde{B}^0$&-&{\bf 55.8\%}&&0.49\%&-\\
			 &$\tilde{W}^0$& 1.6\% &1.0\%&- &{\bf 97.0\%}&0.67\%\\
			 &$\tilde{h}^0$& {\bf 98.3\%} &40.0\%&{\bf 99.5\%}&2.54\%&1.69\%\\
			 &$\tilde{S}^0$&  - &3.20\%& -&-&{\bf 97.4\%}\\\hline
$\tilde{\chi}_3^0$&$\tilde{B}^0$& -& -&-&-&{\bf 95.2\%}\\
			 &$\tilde{W}^0$& 19.8\% &-&-&-&-\\
			 &$\tilde{h}^0$& {\bf 79.8\%} &{\bf 98.9\%}&0.9\%&-&4.05\%\\
			 &$\tilde{S}^0$&   &0.58\%&{\bf 99.1\%}&{\bf 99.98\%}&0.48\%\\\hline
$\tilde{\chi}_4^0$&$\tilde{B}^0$& -  &43.3\%&{\bf 99.6\%}& - &-\\
			 &$\tilde{W}^0$& - &2.31\%&-&-&-\\
			 &$\tilde{h}^0$& &{\bf 53.6\%}&-&{\bf 99.51\%}&{\bf 99.1\%}\\
			 &$\tilde{S}^0$& {\bf  99.8\%} &0.83\%&-&-&0.53\%\\\hline
$\tilde{\chi}_5^0$&$\tilde{B}^0$&{\bf 99.7\%} &-&-&0.54\%&4.52\%\\
			 &$\tilde{W}^0$& - &{\bf 96.3\%}&{\bf 99.3\%}&2.36\%&2.53\%\\
			 &$\tilde{h}^0$&  &3.62\%&0.69\%&{\bf 97.1\%}&{\bf 91.8\%}\\
			 &$\tilde{S}^0$&  -- &-&-&-&1.13\%\\
\hline
\end{tabular}
\end{center}
\end{table}

\subsection{Selecting the renormalisation schemes}
\label{subsec:select-scheme}
As we discussed in detail for the neutralinos, the choice of the renormalisation scheme is crucial for a most
efficient extraction of the counterterms. For instance we argued that $\delta M_1$ will be badly reconstructed if
the bino-like neutralino was not used as an input parameter in a scenario with little mixing. This is the reason we
will adapt the renormalisation scheme  for each benchmark point. We will compare the predictions for the
same observable for different renormalisation schemes within the same benchmark. The difference between these
schemes lies in how we extract the (six) underlying parameters entering the neutralino/chargino sector. For all OS
schemes, the chargino masses are always chosen as input. 
We consider the  following schemes
\begin{itemize}
\item Fully OS schemes with four neutralino masses classified as ${\cal {P}}_6$.  These schemes will be denoted as
\subitem{{\bf --}}  $\textrm{OS}_{1234}$  when taking the 4  lightest neutralinos 
\subitem{{\bf --}} $\textrm{OS}_{2345}$  when taking the 4  heaviest
neutralinos 
\item OS  schemes  where we  take a $\drbar$ condition for $\delta t_\beta$ in addition to three neutralino masses as input. We do so in order to have a better determination of $\tb$ and decouple the system
of equations for the neutralinos and charginos. These schemes will be denoted as
\subitem{{\bf --}} $t_{123}$ when taking the 3 lightest neutralinos
\subitem{{\bf --}} $t_{345}$ when taking the 3 heaviest neutralinos
\subitem{{\bf --}} $t_{134}$ when taking the first, third and fourth
neutralinos
\item Fully OS schemes of the ${\cal {P}}_8$ class where some masses from the
Higgs sector are used as inputs. To fully determine the system we need all in all (including the
chargino masses) eight  input parameters in this case. We resort to these schemes since as pointed earlier schemes based on using solely the masses of the neutralinos are not expected to be good
enough in reconstructing neither $t_\beta$ nor $\lambda$. These two parameters will have a strong impact on the couplings of the neutralinos and hence a crucial influence on many of their decays.
In this category we use two types of schemes,
\subitem{{\bf --}}  $OS_{ijkh_2A_1A_2}$ or $OS_{ijkH^+A_1A_2}$ schemes where  three neutralinos, both pseudoscalars Higgses and either $h_2$ or $H^+$ are chosen as
input in addition to the two charginos. The indices $i,j,k$ indicate the relevant neutralinos. For each of these scenarios we avoid taking the mass of the wino-dominated neutralino as input since $M_2$ is well extracted form the chargino mass measurements. 
\subitem{{\bf --}} $OS_{ijh_2H^+A_1A_2}$ schemes where  two neutralino masses  as well as the Higgs singlet, charged Higgs and
the two pseudoscalar Higgses are used as inputs. 
\item Full $\drbar$ scheme is also used for comparison. In this case we
take the renormalisation scale $\bar{\mu}$ at the mass of the decaying particle as discussed earlier.
\end{itemize}

\noi For processes involving sfermions decays we stick with only one scheme as described in section~\ref{sec:ren_sf}.

\section{One-loop results for neutralino masses and neutralino/chargino decays to gauge bosons}
\label{sec:one-loop-results-chi}
In the absence of not too large mixings between the different components in the
$\tilde{\chi}^0/\tilde{\chi}^\pm$ sector like in the 5 points we have chosen,
the masses of the physical states are determined essentially by $M_1,M_2,\mu, 2
\kappa s$. These parameters must therefore be determined accurately for a
precise determination of the physical masses. Small contributions to these
masses involve a knowledge of $\lambda$ and $\tb$, but as argued previously the
dependence in these two parameters is expected to be mild. When it comes to the
decays, the situation is different since most decays  involve transitions
between different gauge eigenstates and therefore the decays are very often
quite sensitive on the parameters that set the mixing. Therefore in the decays
we will be more careful about how $\lambda$ and $\tb$ are defined. 

\subsection{Neutralino masses}
\label{subsec:chi-masses}
The calculation of the one-loop corrected neutralino masses only calls for the computation of two-point functions.
Yet, ultraviolet finiteness of the full one-loop corrected neutralino masses is a non trivial check on the theoretical consistency of our set-up and its good implementation in our  automated calculator {\tt SloopS} since a large number of counterterms are involved. 
Depending on the schemes we will select, only one or two neutralino masses  receive corrections at one-loop. For all five points,  we compare the results of the schemes
$OS_{1234}$, $OS_{2345}$, $t_{123}$, $t_{234}$ and $\drbar$ for the masses. Predictions on the masses based on the schemes that rely on the Higgs  sector will be briefly commented upon when we discuss the decays, this is motivated by the fact that these Higgs schemes bring in improvements on the  mixings (essentially $\lambda$ and to a lesser degree $\tb$) which are not supposed to be very important for the calculations of the masses. 
\begin{table}[!h]
\caption{\label{tab:results}One-loop corrected masses of neutralinos for different schemes and benchmark points. In bold,
points for which the masses cannot be computed reliably.
All masses are given in GeV. The one-loop corrections for all five neutralino masses in the $\drbar$ scheme  are also given.}
\begin{center}
\begin{tabular}{|c|c|c|c|c|c|c|}
\hline
Scheme&Masses&Point 1 & Point 2 & Point 3 & Point 4 & Point 5\\
\hline
\multirow{2}{1cm}{$OS_{1234}$}&$m_{\tilde{\chi}_5}^{\mathsmaller {tree}}$&1002.17&614.78&1006.64&612.62&574.10\\
&$m_{\tilde{\chi}_5}^{\mathsmaller {1-loop}}$&{\bf 729.01}&614.81&1006.56&608.83&573.22\\
\hline
\multirow{2}{1cm}{$OS_{2345}$}&$m_{\tilde{\chi}_1}^{\mathsmaller {tree}}$&125.67&123.42&112.77&138.09&139.37\\
&$m_{\tilde{\chi}_1}^{\mathsmaller {1-loop}}$&125.56&{\bf -89.66}&{\bf 147.38}&{\bf 205.31}&139.36\\
\hline
\multirow{4}{1cm}{$t_{123}$}&$m_{\tilde{\chi}_4}^{\mathsmaller {tree}}$&500.78&271.67&702.82&603.84&557.31\\
&$m_{\tilde{\chi}_4}^{\mathsmaller {1-loop}}$&{\bf -515.19}&275.13&{\bf 3802.01}&601.19&556.98\\
\cline{2-7}
&$m_{\tilde{\chi}_5}^{\mathsmaller {tree}}$&1002.17&614.78&1006.64&612.62&574.10\\
&$m_{\tilde{\chi}_5}^{\mathsmaller {1-loop}}$&{\bf 1426.14}&614.84&1006.99&613.34&577.17\\
\hline
\multirow{4}{1cm}{$t_{345}$}&$m_{\tilde{\chi}_1}^{\mathsmaller {tree}}$&125.67&123.42&112.77&138.09&139.37\\
&$m_{\tilde{\chi}_1}^{\mathsmaller {1-loop}}$&125.61&{\bf -1808.40}&{\bf -2151.84}&{\bf -479.85}&138.54\\
\cline{2-7}
&$m_{\tilde{\chi}_2}^{\mathsmaller {tree}}$&257.30&200.86&123.80&193.12&276.19\\
&$m_{\tilde{\chi}_2}^{\mathsmaller {1-loop}}$&257.83&{\bf 146.03}&{\bf 1236.66}&189.51&{\bf 74.11}\\
\hline
\multirow{4}{1cm}{$\drbar$}
&$m_{\tilde{\chi}_1}^{\mathsmaller {1-loop}}$& 136.00& 124.10&120.38 &140.90 &147.84 \\
\cline{2-7}
&$m_{\tilde{\chi}_2}^{\mathsmaller {1-loop}}$&265.61  &204.60  &129.52  &  204.23&278.56\\
\cline{2-7}
&$m_{\tilde{\chi}_3}^{\mathsmaller {1-loop}}$& 286.68 &259.56  &241.533  & 280.01 &395.25\\
\cline{2-7}
&$m_{\tilde{\chi}_4}^{\mathsmaller {1-loop}}$&500.72  &  278.72& 703.09 & 601.75 &557.64\\
\cline{2-7}
&$m_{\tilde{\chi}_5}^{\mathsmaller {1-loop}}$&995.41  &  625.50& 1009.26 &  613.84&577.47\\
\hline
\end{tabular}
\end{center}
\end{table}

\noi We advocated that a good scheme should include at least one bino-like, one singlino-like and one higgsino-like from the neutralino sector (the wino-like being well reconstructed from the chargino masses). The numerical results given in
Table~\ref{tab:results} generally follow our expectations. \\
\noi In the $OS_{1234}$ scheme, the only mass to be predicted is that of the heaviest neutralino,
$m_{\tilde{\chi}_5^0}$.
For Point 1 the latter is dominantly bino. Since $M_1$ can not be reliably extracted from the four input masses 
$m_{\tilde{\chi}_{1,2,3,4}^0}$, the corrected mass $m_{\tilde{\chi}_5^0}$ is not trustworthy giving a correction of about 30\%. This is in contrast with Point 2 and
3 where the heaviest neutralino is dominantly wino. In this case the chargino masses constrain $M_2$ very well. In both cases
the heaviest neutralino receives a mass correction at the per-mil level. A similar statement can be made for Points 4 and 5 for which the heaviest neutralino is a higgsino whose main parameter, $\mu$,  is quite well constrained by the input from
the charginos. Note that the correction here, though very modest,  is slightly larger than in the case of the wino due to the fact that a
full reconstruction still requires a knowledge of the underlying $\tb$ and even $\lambda$ for Point 5. \\
\noi In the $OS_{2345}$ scheme, the only mass to be predicted is that of the lightest neutralino,
$m_{\tilde{\chi}_1^0}$. As before the masses which get the smallest correction
correspond to the wino-like LSP, Point 1
and Point 5. This is in sharp contrast to the singlino in Point 2 and the bino in Point 4 whose masses receive very
large corrections. The LSP in Point 3, although higgsino-like, gets a
non-negligible correction. This means, a point we hinted at previously, that the
chargino system does not fully define the higgsino-like neutralino due to the
reconstruction of $t_\beta$.
\noi We expect the scheme $t_{123}$ to fare better than $OS_{1234}$.
Indeed, as compared to $OS_{1234}$ the masses of the heaviest neutralino are changed very little for Points 2,3
and 5 and are predicted with a smaller correction for Point 4. The  $t_{123}$ scheme however does not improve the situation for Point 1 where the mass of the heaviest bino is as always badly reconstructed. The same problem afflicts the prediction of the mass of
the bino-like $\chi^0_4$ for Point 3 and of the singlino for Point 1. Otherwise the corrections for $\neutf$ are
negligible since these neutralinos are either winos or higgsinos.\\
\noi Similar arguments explain the results for the one-loop corrected neutralino masses in the scheme $t_{345}$.
This scheme works well for Point 1 as the three heaviest neutralinos correspond to the bino, singlino and dominantly higgsino. For Point 2, the singlino component is not accessed which explains why the mass of the dominantly singlino $\neuto$ cannot be predicted reliably. The correction to the mass of $\neutt$, a dominantly bino neutralino with a
large higgsino admixture, is also large, around 30\%. For Point 3, the two lightest neutralinos are higgsino-like
and receive very large corrections, this illustrates the futility in
using the mass of the neutral wino as input at the expense of one of the other neutralino masses. For Point 4 the wino dominated
neutralino receives small corrections while, as expected, the mass of the dominantly bino $\neuto$ is unreliable.
Similarly for Point 5 where the singlino dominated $\neutt$ is unreliably predicted.
Note that a scheme which does not allow a good reconstruction of some of the
parameters, for example the singlino mass term $\kappa s$, can be nevertheless
appropriate for observables where the singlino component does not play a role.

\noi It is also interesting to look at the predictions given by a $\drbar$ scheme, the renormalisation scale
$\bar{\mu}$ is taken at the (tree-level) mass of the particle. Here corrections to all masses are
calculated. For all masses and for all points, the corrections are small and never exceed 10\%, compare Table~\ref{tab:bench} and Table~\ref{tab:results}. However, note that
when the underlying parameters for the OS scheme are reconstructed efficiently,  the OS scheme for that particular
mass gives smaller corrections than the $\drbar$ scheme.

\noi To summarise,  in order to compute radiative corrections to the masses reliably, one then has to be careful about
the choice of the renormalisation scheme. A good scheme should be chosen according to the characteristics of the
point considered in that the input parameters should reconstruct the main ingredients that define the nature of the
particle whose mass is to be corrected at one-loop. The $\drbar$ scheme is versatile and reliable but a {\it good} OS scheme fares
better, in the sense of leading to smaller corrections,  as far as masses are concerned.

\subsection{Two-body neutralino/chargino decays to  a gauge boson}
\label{subsec:chi-dec}
We now study the one-loop corrections to decays of the type $\chi_i \to \chi_j^\prime V, V=W^\pm,Z$. 
If the charginos and neutralinos did not mix these  transitions would not be possible at all. This of course applies
to a (pure) singlino state. It also applies to transitions between two neutralinos through a $Z$ for the case of a wino and
bino. This transition is only possible among higgsinos but there is generally little mass difference between these higgsinos for 
these decays to occur  on-shell. Other transitions are possible among winos and separately among
higgsinos, but again phase-space is restrictive. These observations together with those we made about mixing in section~\ref{subsec:choiceP} explain the
main features of the decays. We will study
some of the schemes we used for the calculations of the masses. For each of the five points we will add another
scheme of the category ${\cal {P}}_8$ that requires inputs from the Higgs sector. We restrict ourselves to processes
that have a branching ratio at tree-level of at least 1\% since they are the only ones of any physical relevance.
 In the $\drbar$ scheme, we take the scale $\bar{\mu}$ at the mass of the decaying particle.

\subsubsection{Point 1}
\label{sec:gaugedecay_point1}
\begin{table}[!htb]
\caption{Point 1 : Partial widths (in MeV) for decays of neutralinos and charginos into one gauge boson at
tree-level (tree) and at one-loop (tree + one-loop) with four different renormalisation schemes. The
relative correction to the partial decay widths  is also indicated in parentheses. The schemes for the one-loop results (tree + one-loop), here $t_{234}$,  $OS_{2345}$, $OS_{245h_2A_1A_2}$ and $\drbar$, are defined in the text.
}
\begin{center}
\label{tab:point1}
\begin{tabular}{llllll}\hline 
& tree & $t_{345}$&$OS_{2345}$ &$OS_{245h_2A_1A_2}$ & $\drbar$ \\\hline 
$\chargt\to W^+\neuto$  \;\;\; &  $406  $ & $412 $   (1\%)     &   $419 $ (3\%)  &  $420  $ (3\%) & $417$ (3\%)\\
$\chargt\to Z\charg$   &   $341  $&$349  $ (2\%)  &$357 $ (5\%)&   $355  $  (4\%)&  $354  $  (4\%)\\\hline
$\neutt\to W^-\charg$   &  $271   $& $274 $ (1\%)  &$280  $ (3\%) &  $280$  (3\%) & $276$ (2\%) \\
 $\neutt\to Z\neuto$  &$183    $& $184  $ (0.8\%)    & $192  $ (5\%) &  $190    $ (4\%) &$190    $ (4\%) \\\hline
 $\neuth\to W^-\charg$ &   $452  $  & $456   $ (0.9\%) &$467$ (3\%) &   $461  $ (2\%) &  $458$ (1\%) \\
 $\neuth\to Z\neuto$  & $33.5  $ &  $37.2  $ (11\%)   &  $33.8  $ (1\%) & $35.1  $ (5\%)  
& $30.2$ (-10\%) \\\hline
 $\neutf\to W^-\charg$  &  $10.4 $ &  $10.6  $  (2\%) &  $18.2  $ (75\%) &  $9.56 $ (-8\%) & $9.54 $ (-8\%) \\
 $\neutf\to W^-\chargt$  & $22.9  $ &  $26.3  $  (15\%) &  $42.1  $  (84\%) &
$23.2  $ (1\%) & $24.6 $ (7\%)\\
 $\neutf\to Z\neuto$   & $6.26  $ &  $6.44  $   (3\%)&  $11.0  $  (76\%)  & $5.83  $  (-7\%) & $5.70  $  (-9\%)\\
 $\neutf\to Z\neutt$    &$26.2  $ &  $29.9  $  (14\%) &  $47.7  $ (82\%)  &$26.1  $ (-0.7\%) &$28.1  $ (7\%)\\
 $\neutf\to Z\neuth$   &$3.12  $ &  $3.64  $ (17\%)  &  $6.02  $ (93\%)  &$3.44  $ (10\%)&$3.16  $ (1\%) \\\hline
 $\neutfi\to W^-\charg$  &$26.8  $ &  $22.4  $  (-14\%) &  $26.8  $( 0.1\%) &$26.1  $ (-2.5\%) & $27.3 $ (2\%)\\
 $\neutfi\to W^-\chargt$ &    $611 $   & $618 $ (1\%)  &  $625 $ (2\%)& 
$624 $ (2\%) & $630$ (3\%) \\	
 $\neutfi\to Z\neutt$   & $515 $  &  $517 $  (0.4\%) &  $533 $  (4\%)  & $531
$ (3\%) & $531 $  (3\%) \\
 $\neutfi\to Z\neuth$ &  $118 $    &   $122 $  (3\%) &  $116 $  (-2\%)  & 
$117 $ (-1\%) &  $121 $ (3\%)\\       \hline  
\end{tabular}
\end{center}
\end{table}

\noi This benchmark point,  see
Table~\ref{tab:compo}, can be characterised  as 
$(\tilde{W}^0_3,\tilde{H}^0,\tilde{H}^0,\tilde{S}^0,\tilde{B}^0)$ and $(\tilde{W}^+,\tilde{H}^+)$  according 
to our discussion in section~\ref{subsec:choiceP}. The LSP neutralino is very much wino-like.
Choosing the LSP mass as an input is redundant since, as we saw, the extraction of $M_2$, which sets the
characteristics of the wino, from the chargino mass is sufficient. Therefore, we take here the $t_{345}$ and
$OS_{2345}$ schemes to get access
to a maximum of information on the neutralino sector. We did expose at some length the shortcomings of these schemes when it comes to a good reconstruction of the singlino and higgsino characteristics and for the need to revert to
mass measurements from the Higgs sector. The latter  should constrain $\lambda$ much
better and contribute to improve the determination of  $\tb$. 
We therefore study the predictions of the scheme $OS_{245h_2A_1A_2}$. Note
that we advocate the use of the
mass of the bino-like and the singlino-like neutralino in conjunction with the Higgs masses. We first observe that
in the scheme $OS_{245h_2A_1A_2}$ the one-loop corrected neutralino masses are
$m_{\neuto}=125.66$ GeV and
$m_{\neuth}=278.97$ GeV. These are per-mil level corrections. 
Therefore {\it \`a priori} this scheme seems to be indeed a very good scheme. This statement is confirmed if one
looks at the one-loop corrections to all the decays listed in Table~\ref{tab:point1}. For all decays not involving the singlino dominated
neutralino $\chi_4^0$ the corrections in the scheme $OS_{245h_2A_1A_2}$ are below 5\%. Even for singlino decays, the
corrections are not larger than
$10\%$, the largest correction is reached for the smallest branching fraction. The corrections in this scheme are
smaller than in the $\drbar$ scheme  for all the considered channels but $\chi^0_4 \to \chi^0_3 Z$ which is the
smallest branching ratio for the singlino dominated $\chi^0_4$. 

\noi Both the $t_{345}$ and $OS_{2345}$ schemes give small
corrections (typically less than $10\%$) for all channels apart those involving the decay of the singlino. For the case of the
singlino decays the corrections in the mixed $\drbar$-OS scheme $t_{345}$ are under control (below $20\%$) but they should
not be trusted in the scheme $OS_{2345}$. To summarise,  a reconstruction of
$\lambda$ is essential 
 to compute the decays of the singlino. As expected neutralino/chargino mass measurements do not allow a good reconstruction in the $OS_{2345}$ scheme while it is perfectly fine when it comes to the decays of the other particles. We would have expected the scheme $t_{123}$ to do as badly as the  $OS_{2345}$ scheme for singlino-dominated decays since a direct
access to $\lambda$ is not possible here also. However, in the $t_{123}$
scheme, $\tb$ is solved independently thus permitting a better access to
$\lambda$, even if not as good as in $OS_{245h_2A_1A_2}$. 

\noi We argued earlier that the mixings in this sector are not very sensitive to $\tb$. To quantify this
statement we looked precisely at the determination of the finite part of each of
the counterterms for $(\mu,\tb, \lambda)$ for the three schemes. We have,
respectively for the schemes $t_{123};OS_{2345};OS_{245h_2A_1A_2}$,  
$$(\delta \mu/\mu, \delta \tb/\tb,
\delta \lambda/\lambda)_{\rm finite}=\overbrace{(-2.3\%,0,+6\%)}^{t_{123}};\overbrace{(-2.6\%, -20\%,+43\%)}^{OS_{2345}};\overbrace{(-2.6\%,-17\%,+0.9\%)}^{OS_{245h_2A_1A_2}}$$ which shows first of all that $\mu$ is well reconstructed, independently of the scheme. This is easy to understand since all
schemes rely on the chargino masses and confirm that the $\tb$ dependence in the masses is very weak. Note that both $OS_{2345}$ and $OS_{245h_2A_1A_2}$ do not determine $\tb$ well. This was to be expected from the discussion in Section~\ref{sec:appli-benchmarks}, {\it i.e}, masses of the neutralinos and Higgses are not much dependent on $\tb$. Using extra measurements
with the Higgs masses improves the extraction of $\tb$ only marginally. Yet, for these decays, the $\tb$ dependence is generally very weak and so is therefore the $\tb$ scheme dependence. What makes a huge difference is the extraction of $\lambda$.  
We see that the $OS_{2345}$ is ineffective, the contribution of the counterterm alone would contribute about $2\times 43\%
\sim 86\%$ in decays directly proportional to $\lambda$ like those in transitions involving $\chi^0_{4}$. Indeed, for $\chi^0_{4}$ decays  the difference between  the $OS_{2345}$ and $OS_{245h_2A_1A_2}$ can be, to a large extent, accounted for by the difference in the value of $\delta \tb$ for all channels listed in the decays of  $\chi^0_{4}$. 

\noi Two final remarks concerning this point. Leaving aside the decays of the
problematic singlino, we see that the differences between the schemes is quite
small, in fact the largest discrepancies occur when the branching fraction is
smallest among all the decays of a given neutralino. Again, the small branching
fraction is an indication of the smallness of the coupling which is most
sensitive to mixing and hence would be most dependent on the scheme. The
corrections  though generally small can not be accounted simply by taking an
effective running of the gauge coupling (which would give a correction of about
$7\%$), there are therefore genuine electroweak corrections. These observations
should be kept in mind for the other points we will study. 

\subsubsection{Point 2}
\label{sec:gaugedecay_point2}
\begin{table}[!htb]
\caption{Point 2: same as Table~\ref{tab:point1}. }
\begin{center}
\label{tab:point2}
\begin{tabular}{llllll}\hline 
 &tree& $t_{123}$&$OS_{1234}$&$OS_{12h_2A_1A_2H^+}$ & $\drbar$ \\\hline
 $\chargt\to W^+\neuto$  \;\;\;  &  79.0   &   155  (96\%)    &  182  (130\%)  & 84.4  (7\%) &  58.5   (-26\%) \\
$\chargt\to W^+\neutt$  \;\;\;  & 368   &   216   (-41\%)   &    134 (-64\%)  &  293 (-20\%) & 266  (-28\%) \\
$\chargt\to W^+\neuth$  \;\;\;  & 1370   & 1200   (-12\%)     &  1140   (-17\%) &  1180 (-14\%)& 1170 (-15\%) \\
$\chargt\to Z\charg$  &1400    &  1270(-9\%)  & 1210 (-14\%) & 1240 (-12\%)  & 1221 
(-13\%) \\\hline 
$\neuth\to Z\neuto$ & $34.5 $ & $ 70.9 $ (106\%) & $ 86.8 $ (152\%) & $ 42.7 $ (24\%) & $ 29.6 $ (-14\%) \\\hline
$\neutf\to Z\neuto$ & $11.3 $ & $ 23.9 $ (110\%) & $ 26.8 $ (137\%) & $13.3 $ (18\%)& $11.9 $ (5\%)\\\hline
  $\neutfi\to W^-\charg$ &   1430  &   1270  (-12\%) &   1210 (-16\%) & 1240 (-14\%) & 1220 (-15\%)\\
  $\neutfi\to Z\neuth$   & 1250  &  1120   (-11\%)&   1040 (-17\%) & 1090 (-13\%) & 1080 (-14\%)\\
 $\neutfi\to Z\neutf$  &  58.8    &   55.1 (-6\%) &  65.1  (11\%) &  60.4 (3\%) &  57.3 (-2.5\%) \\       \hline  
\end{tabular}
\end{center}
\end{table}
\noi This benchmark point features a singlino LSP where the neutralinos can be characterised as
$(\tilde{S}^0,\tilde{B}^0,\tilde{H}^0,\tilde{H}^0,\tilde{W}^0_3)$ and the charginos
$(\charg,\chargt)=(\tilde{H}^+,\tilde{W}^+)$. $\neutt$ and $\neutf$ are almost equal mixture of bino and higgsino. Note
that phase space does not allow the decay $\neutt \to Z\neuto $. Due to the
singlino nature of the LSP, decays to the LSP $\neuto$ have a very small partial
width. The only exception is $\neutfi \to Z\neutf$, however this is due to
the very small higgsino
component  in $\neutf$  compared for example to  $\neuth$ which is almost pure higgsino (recall that the coupling
of the $Z$ to neutralinos requires both neutralinos to be
higgsino-like). The need to access the singlino and
bino
component leads us to take the masses of their corresponding neutralinos as input. We therefore consider the schemes $t_{123}$ and $OS_{1234}$. As argued previously this does not guarantee a good reconstruction of the mixing
$\lambda$ which is crucial
in calculating the decays where the singlino dominated state, $\neuto$, is involved.

\noi For the OS scheme that uses the Higgs mass measurements we advocate $OS_{12h_2A_1A_2H^+}$ where only the singlino-like and bino-like are used from the neutralino sector (we still of course use the masses of the charginos) as well as four Higgs masses, the next-to-lightest CP-even neutral Higgs, the two CP-odd neutrals and
the charged Higgs.  The aim here is a better determination of $\lambda$. We recall first that this scheme does a very
good job in predicting the masses of the three
heaviest neutralinos with corrections below 1\% for all three masses (we find for the corrected masses, $m_{\chi^{0}_{2,3,4}}=253.12, 273.63, 614.82$ GeV). 
Table~\ref{tab:point2} shows clearly that the $OS_{12h_2A_1A_2H^+}$ and $\drbar$ schemes are the
ones that give the smallest corrections to the decays especially when the singlino is involved. In many channels the predictions of the full $\drbar$ are
within a few per-cent of what is obtained with the $OS_{12h_2A_1A_2H^+}$ scheme, for these decays the impact of
singlino mixing is marginal. It is therefore not surprising that for these same channels the prediction of
the other two schemes agree within 5\%. The channels where the difference between the $\drbar$ and
$OS_{12h_2A_1A_2H^+}$ scheme is above 5\% are those where the $t_{123}$ and the $OS_{1234}$ become totally
unreliable with corrections of order $100\%$ or worse. Not surprisingly the worst cases involve decays into a
singlino, channels where the partial widths are very small. These channels require an excellent knowledge of the
full mixing structure of the singlino. To confirm these findings we have extracted, as for Point 1, the finite part
of each of the counterterms for $(\mu,\tb, \lambda)$ for the three schemes $(t_{123};OS_{1234};
OS_{12h_2A_1A_2H^+})$, 
we find 
$$(\delta \mu/\mu, \delta \tb/\tb,
\delta \lambda/\lambda)_{\rm finite}=\overbrace{(-2.6\%,0,57\%)}^{t_{123}};\overbrace{(-3.0\%, 23\%,+73\%)}^{OS_{1234}};\overbrace{(-2.8\%,12\%,+10\%)}^{OS_{12h_2A_1A_2H^+}}.$$
These values confirm the
observation that we made for Point 1, $t_{123}$ and $OS_{2345}$ schemes entail large corrections for $\delta
\lambda/\lambda$. We note that for decays into singlinos the difference between the relative corrections given by the schemes can be approximated by $\Delta(\delta \Gamma/\Gamma) \sim \Delta (\delta \lambda/\lambda)$. It should also be observed that although $\tb$ is not an issue for these decays, the
finite part of $\tb$ in the $OS_{12h_2A_1A_2H^+}$ is not so small, it amounts to about $12\%$.
We have also computed one-loop
corrections in a scheme where we rather chose the lightest CP-even Higgs $OS_{245h_1A_1A_2}$ and we found
similar corrections for the MSSM-like transitions but large corrections for singlino processes. Once again using the singlet Higgs $h_2$ allows for a better reconstruction of $\delta\lambda$.

\subsubsection{Point 3}
\label{sec:gaugedecay_poin3}
\begin{table}[!htb]
\caption{Point 3: same as Table~\ref{tab:point1}.
}
\begin{center}
\begin{tabular}{llllll}\hline 
&tree& $t_{134}$ & $OS_{1234}$ & $OS_{34h_2A_1A_2H^+}$ & $\drbar$\\\hline $\chargt\to W^+\neuto$ \;\;\;& 1960 & 1670(-15\%) & 1830 (-7\%) & 1680 (-15\%) & 1670 (-15\%) \\
$\chargt\to W^+\neutt$ \;\;\;& 2110 & 1730 (-18\%) & 1870 (-11\%) & 1740 (-17\%) & 1730 (-18\%) \\
$\chargt\to Z\charg$&   2050  &  1710 (-16\%)  & 1860 (-9\%) & 1720 (-16\%) & 1730 (-16\%)  \\\hline
 $\neuth\to W^-\charg$ &     11.2   & 20.5 (84\%)  &  27.0  (141\%) & 12.0 (7\%) & 12.0 (7\%)\\
  $\neuth\to   Z\neuto$ &   $ 4.32 $ & $ 7.68 $   (78\%) &  11.7   (171\%) &  4.51  (4\%)  & $ 4.20 $ (-3\%)\\
$\neuth\to Z\neutt$ & $ 3.78 $ & $ 7.00 $ (85\%) & $ 8.34 $ (120\%) & $ 4.08 $ (9\%) &$ 4.28 $ (13\%) \\\hline
$\neutf\to W^-\charg$ &    445 & 479 (8\%)  & 525 (18\%) & 480 (8\%) & 460 (3\%)\\
 $\neutf\to Z\neuto$ &  $ 99.4$ & 113 (13\%) & 70.3  (-29\%)  & 110  (11\%)  &  105  (6\%) \\
 $\neutf\to Z\neutt$ &   306 &  328  (7\%) &  410  (34\%) &   332 (8\%)&  323 (6\%)  \\\hline
 $\neutfi\to W^-\charg$ &    2060  & 1720 (-16\%) &  1860      (-10\%) & 1720 (-16\%) & 1710 (-17\%)\\
  $\neutfi\to Z\neuto$ &  578 &  500   (-14\%)  &  256   (-56\%) &  487(-16\%)  &  502 (-13\%)  \\
 $\neutfi\to Z\neutt$ &   1480    &  1220 (-17\%) & 1620 (9\%) & 1240 (-16\%)  & 1270 (-13\%)  \\       \hline  
\end{tabular}
\label{tab:point3bis}
\end{center}
\end{table}

\noi Once the identification of the neutralinos has been made we realise that
this point is somehow a reshuffling of Point 2. The r\^ole played by $\neuto$ is
replaced by $\neuth$ and the neutral higgsino-like have become $\neuto,
\neutt$. 
Therefore we expect similar conclusions to emerge, especially as regards singlino-like decays,  even though  all the states are almost pure here. Following our
characterisation, the point is identified as $(\tilde{H}^0,\tilde{H}^0,\tilde{S}^0,\tilde{B}^0,\tilde{W}^0_3)$ and
$(\tilde{H}^+,\tilde{W}^+)$. The hierarchy in the charginos has not changed as compared to Point 2. We therefore advocate the use of the scheme
$OS_{34h_2A_1A_2H^+}$. A careful look at the predictions of the different
decays, taking into account the identity of the particles involved in the decay,
shows similar corrections as those for Point 2 and whenever large
discrepancies occur they can be explained along the same arguments as those we have just put forward for Point 2, in
particular the schemes $t_{134}$ and $OS_{1234}$ give unreliable predictions when the decay involves the singlino-like
$\neuth$. \\
\noi We should point at another issue not directly related to the singlino. For the decays  $\neutf, \neutfi \to \neutt Z$
the predictions in the scheme $OS_{1234}$ differ by  $+27\%$ with  those in the scheme $t_{134}$, 
while  the difference between $OS_{34h_2A_1A_2H^+}$ and
$t_{123}$ is just $+1\%$. Similarly for $\neutf, \neutfi \to \neuto Z $ the difference is $-42\%$ (between the $OS_{1234}$ and 
$t_{134}$ schemes) and about $-2\%$ (between $OS_{34h_2A_1A_2H^+}$ and $t_{123}$).
This has to do with the $\tb$
dependence. Although these decays are mildly sensitive to $\tb$, it remains that $\tb$ is so badly reconstructed
in the scheme $OS_{1234}$ that it leads to noticeable differences with the prediction for the other schemes. 
To wit, if we again look into the finite part of the counterterms we find
$$(\delta \mu/\mu, \delta \tb/\tb,
\delta \lambda/\lambda)_{\rm finite}=\overbrace{(-3.3\%,0,42\%)}^{t_{123}};\overbrace{(-1.5\%, -169\%,80\%)}^{OS_{1234}};\overbrace{(-3.2\%,-5.39\%,+4\%)}^{OS_{12h_2A_1A_2H^+}}.$$ We see that the scheme
$OS_{1234}$ fares quite badly ($\delta \tb/\tb |_{\rm finite} \sim -170\%$). These values also confirm the
observation that we made for Point 1, $t_{123}$ and $OS_{1234}$ schemes entail large corrections for $\delta
\lambda/\lambda$. Note also that here $OS_{12h_2A_1A_2H^+}$
proves to be a quite good scheme, in particular for both $\tb$ and $\lambda$. The fact that the extraction of $\tb$
proves very uncertain for this point  in the $OS_{1234}$ scheme  is easy to understand. Remember that all charginos and neutralinos are to a very good approximation almost in a pure state, therefore from their masses the small $\tb$ dependence hidden in the
mixing is reconstructed badly (inversely proportional to the very small mixing).

\subsubsection{Point 4}
\label{sec:gaugedecay_point4}
\begin{table}[!htb]
\caption{Point 4: same as Table~\ref{tab:point1} for point 4.  }
\begin{center}
\begin{tabular}{llllll}\hline 
&tree& $t_{123}$&$OS_{1234}$ & $OS_{134A_1A_2H^+}$ & $\drbar$ \\\hline 
$\chargt\to W^+\neuto$  \;\;\;&     307  & 364  (19\%)       & 388     (26\%) & 371 (21\%) & 379 (23\%) \\
$\chargt\to W^+\neutt$  \;\;\; &  1420  & 1340     (-6\%)    & 1040    (-27\%) &  1220 (-14\%) &1360 (-4\%) \\
$\chargt\to Z\charg$  & 1300   & 1210  (-7\%)  &950   (-27\%) & 1160 
(-10\%)&1260(-3\%)\\\hline
   $\neutf\to W^-\charg$    &  1310 &1210  (-7\%)    & 960  (-27\%) & 1160 (-11\%) & 1260 (-3\%)\\
 $\neutf\to Z\neuto$   & 383 &  425  (11\%)  &  232  (-40\%) &  413  (8\%)  &  417  (9\%) \\
 $\neutf\to Z\neutt$    & 1020  & 916   (-10\%)& 605  (-41\%) & 927 (-9\%) & 1000 (-2\%) \\\hline
 $\neutfi\to W^-\charg$  &  1340  & 1240   (-8\%)  & 1050 (-22\%) & 1150  (-14\%) & 1230 (-8\%)\\
  $\neutfi\to Z\neuto$   & 89.5& $109 $  (21\%) &  $152$ (70\%) & 110 (23\%) & 103 (16\%)\\
 $\neutfi\to Z\neutt$  & 165   &  169  (2\%) &  293 (77\%) & 165(-0.6\%) &
158(-4\%) \\       \hline  
\end{tabular}
\label{tab:point4}
\end{center}
\end{table}
\noi The most important feature of this scenario is that the singlino is for all purposes totally decoupled, here
$\lambda$ is extremely small. In particular, the near-pure singlino nature of $\neuth$ explains why its partial decay
width into
gauge bosons is strongly suppressed, the largest two-body decay involving a gauge boson is into $W\charg$ with a
partial width $\Gamma =2.09\times 10^{-5}$ GeV. The preferred decays of 
$\neuth$ involve higgses which we do not study here. We
should therefore not be surprised that in Table~\ref{tab:point4} $\neuth$ is not present. This benchmark point is characterised as
$(\tilde{B}^0,\tilde{W}^0_3,\tilde{S}^0,\tilde{H}^0,\tilde{H}^0)$; $(\tilde{W}^+,\tilde{H}^+)$ where the LSP is
bino-like. Therefore  the mass of the LSP features in all the OS schemes. For the scheme
that relies on inputs from the Higgs sector we propose  $OS_{134A_1A_2H^+}$. Since we are looking at decays in
what is essentially the MSSM where $M_{1,2},\mu$ are well reconstructed, any discrepancy between the schemes has to do
with $\tb$. We first observe that the fully $\drbar$ scheme {\it and}
the $t_{123}$ scheme give predictions which for all decays shown in ~Table.~\ref{tab:point4} are within 5\%, the
only exception
is $\neutf\to Z\neutt$ where the difference is $8\%$. The $OS_{134A_1A_2H^+}$ scheme also agrees with the fully
$\drbar$ scheme within $10\%$. For a few decays the $OS_{1234}$ scheme gives
corrections of order $70\%$ and $-40\%$ that are quite different from the results in the other schemes. These
observations lead to suspect that once again the determination of $\tb$ is in question especially in the $OS_{1234}$ scheme.  Indeed if we look at the finite part of the counterterms we find $$(\delta \mu/\mu, \delta \tb/\tb,
\delta \lambda/\lambda)_{\rm finite}=\overbrace{(0.03\%,0,-1430\%)}^{t_{123}};\overbrace{(0.3\%, +150\%,-4500\%)}^{OS_{1234}};\overbrace{(0.08\%,27.4\%,2841\%)}^{OS_{134A_1A_2H^+}}$$
showing the very poor reconstruction of $\tb$ in $OS_{1234}$ and to a lesser extent in $OS_{134A_1A_2H^+}$.

\noi Observe that a good reconstruction of $\tb$ has also an impact on the reconstruction of $\mu$, although all three schemes perform well for $\mu$,
 $t_{123}$ does ten times better than $OS_{1234}$. Since many decays into gauge bosons are triggered from
higgsino to higgsino transitions a very precise determination of $\mu$ is important. Note in passing that
$\delta \lambda$ is totally unreliable as expected since the resolution of the system leads to a division by $\lambda$,
{\it i.e.} a division by a very small number. This has no direct impact on the corrections computed since we did not consider decays involving singlinos for this point.
To summarise for this point all schemes, apart from $OS_{1234}$, do a good job giving moderate corrections. The
largest correction of order 20\% occurs for $\chargt\to W^+\neuto$. This is a genuine correction which is fairly
independent of the scheme. 

\subsubsection{Point 5}
\label{sec:gaugedecay_point5}
\begin{table}[!htb]
\caption{Point 5: same as Table~\ref{tab:point1}. }
\begin{center}
\label{tab:point5}
\begin{tabular}{llllll}\hline 
 &tree& $t_{123}$&$OS_{2345}$  & $OS_{234A_1A_2H^+}$ & $\drbar$ \\\hline 
$\chargt\to W^+\neuto$  \;\;\; & 1250   & 1160  (-7\%)        & 909    (-27\%)  & 1150 (-8\%) & 1190 (-5\%) \\
$\chargt\to W^+\neutt$  \;\;\; &531    & 483      (-9\%)   & 545    (3\%)   & 518 (-2\%) & 501 (-6\%)\\
$\chargt\to W^+\neuth$  \;\;\; & 250   & 265     (6\%)    &172     (-31\%)    & 261 (5\%) & 264 (6\%)\\
$\chargt\to Z\charg$  & 1310   & 1220   (-7\%) & 945  (-28\%) & 1210 (-7\%)  & 1240 (-5\%)\\\hline
$\neutt\to W^-\charg$ &    $ 34.3$ &  30.6  (-11\%)  &  37.1  (8\%) & 33.0 (-4\%) & 34.1 (-1\%) \\\hline
 $\neuth\to W^-\charg$ &     58.8 &  55.7   (-5\%) &  $ -17.0 $ (-128\%)  & $ 53.7$ (-9\%) & 57.8 (-2\%)\\ 
 $\neuth\to Z\neuto$  & $ 2.75$ &  2.94   (7\%) &  $5.41 $  (96\%)&  $ 2.97$ (8\%) &$ 2.83$ (3\%)\\\hline
  $\neutf\to W^-\charg$    &1320 & 1220   (-8\%)  & 953  (-28\%)  & 1201
(-9\%) & 
1210(-8\%) \\
 $\neutf\to Z\neuto$  & 1280 &  1160   (-9\%) &  746  (-42\%) & 1150 (-10\%) & 1210 
(-5\%)\\
 $\neutf\to Z\neutt$   &233 &  223  (-4\%)  &  377 (62\%)& 240 (3\%)  & 230(-3\%)\\
$\neutf\to Z\neuth$   &157 &  168  (7\%) &  116  (-26\%) & 165 (5\%)  & 166 (6\%) \\\hline
 $\neutfi\to W^-\charg$  & 1230   & 1120    (-9\%) & 940  (-23\%)   & 1110 (-10\%) & 1150 (-7\%)\\
  $\neutfi\to Z\neuto$   & 108&  106  (-3\%)& 254  (133\%) & 107 (-2\%)  & 101 (-7\%)\\
 $\neutfi\to Z\neutt$ &  166   &   147(-11\%)  &  13.0 (-92\%)  & 155 (-7\%)& 151 (-9\%) \\       \hline  
\end{tabular}
\end{center}
\end{table}
\noi While Point 4 had the smallest $\lambda$ and featured the most decoupled singlino of our benchmarks, Point 5 has the
largest $\lambda$ while the $\mu$ parameter has slightly changed. $\lambda$ is
rather small  but it is large enough to
allow for a few per-cent mixing of the singlino with the higgsino, see Table~\ref{tab:compo}, and subsequently with the other neutralinos, leading for
example to a partial width of $10^{-2}\,{\rm GeV}$ for the decay of the
singlino-like neutralino, $\neutt \to W\charg$.
The LSP is wino-like and the point is characterised as
$(\tilde{W}^0_3,\tilde{S}^0,\tilde{B}^0,\tilde{H}^0,\tilde{H}^0)$ and $(\tilde{W}^+,\tilde{H}^+)$. This hierarchy
suggests to choose the masses of $\neutt$ and $\neuth$ as inputs for all OS schemes. With $t_{123}$ and $OS_{2345}$ we also consider the  $OS_{234A_1A_2H^+}$ scheme where the Higgs masses
are used for a better
extraction of $\delta t_\beta$ and $\delta\lambda$. In this scheme the corrections to the masses of $m_{\neuto}$ and
$m_{\neutf}$ are totally negligible, not exceeding 0.5 per-mil. Moreover the corrections to all the decays we
have considered are quite moderate, below 10\%. Results in the $\drbar$ are very similar, the
difference between the two never exceeds more than $5\%$. The differences between the $OS_{234A_1A_2H^+}$ and
$t_{123}$ are also quite small, within a margin of 7\% even for the decay of the singlino. One can already guess
that the presence of a non negligible $\lambda$ can help not only in better reconstructing this parameter from the
Higgs masses but also in better reconstructing $\tb$. This is in contrast  with the results obtained with the $OS_{2345}$
scheme. The latter is totally  unreliable essentially due to a failed reconstruction of both $\lambda$, see
the large corrections involving the singlino $\neutt$, but also due to a quite bad reconstruction of $\tb$
(see the decays of the other neutralinos in particular the decays of the bino-like $\neuth$). These observations are
borne out by the values of the finite part of the key counterterms in the schemes
$t_{123}, OS_{2345}$ and $OS_{234A_1A_2H^+}$ with  
$$(\delta \mu/\mu, \delta \tb/\tb,
\delta \lambda/\lambda)_{\rm finite}=\overbrace{(-0.1\%,0,-1.2\%)}^{t_{123}};\overbrace{(0.2\%, +115\%,-30\%)}^{OS_{2345}};\overbrace{(-0.1\%,1.9\%,1.6\%)}^{OS_{234A_1A_2H^+}}.$$


\section{One-loop corrections to two-body sfermion decays to fermions}
\label{sec:one-loop-sfermion}
We now compute the one-loop corrections to the decays of third generation sfermions
into a fermion and a neutralino/chargino. These processes are often the preferred decay modes of sfermions and are
the main channels used for third generation squark searches at the LHC
~\cite{Aad:2015pfx,Chatrchyan:2013xna,Khachatryan:2015wza}. Other decay channels involving Higgses will be
considered in a separate publication~\cite{renormalization_higgs}.
For squarks we compute both QCD and EW corrections. As before we include only the decays for which the tree-level
branching ratio is above a few percent as they are the only physically relevant ones.\\
For the definition of the parameters of the sfermions we will consider the scheme presented in section~
\ref{sec:ren_sf}. Namely, for
the squarks the input parameters for the third generation will be $m_{\tilde{b}_1}$, $m_{\tilde{b}_2}$,
$m_{\tilde{t}_1}$ $\theta_{b}$ and $\theta_{t}$ for the squarks and  $m_{\tilde{\tau}_1}$,
$m_{\tilde{\tau}_2}$ and $m_{\tilde{\nu}_\tau}$ for the staus. For the QCD corrections we take $\alpha_s(1 \rm{TeV})=0.0894$,
this scale of $\alpha_s$ corresponds to the mass of $\tilde b_1$ in all of our benchmarks. As in the previous
section we will test different schemes for the neutralino sector. The difference between the latter schemes will impact the predictions for the electroweak corrections. QCD corrections do not impact these schemes but only the squark sector. Since all the OS schemes adopt the same definition for the input parameters for the squark sector there will be no difference between the OS schemes, including the $t_{123}$-type schemes. There may be differences in the QCD corrections between the OS schemes and the full $\drbar$ scheme. As we will see, the QCD scheme dependence is very small and generally hardly noticeable. 

\noi The couplings of the type $\tilde{f} f^\prime \tilde{\chi}$ (for both charged and neutral $\tilde{\chi}$)
responsible for these decays originate from two sources. First, gauge type coulings ($\propto g, g^\prime$) occur
with wino
and bino-like $\tilde{\chi}$. A right-handed $\tilde{f}_R$ will only couple to the bino component. Second, Yukawa
type
couplings 
$$y_{d,u}=(\frac{g\sqrt{2}}{ M_W}) (\frac{m_d}{c_\beta}, \frac{m_u}{s_\beta})  \sim \frac{g\sqrt{2}}{M_W} (m_d \tb, m_u) \; {\rm for \;} \tb > 3$$ 
are important only for third family sfermions in
particular the stops and sbottoms. For sbottoms this coupling is enhanced by $\tb$. Therefore phase-space
allowing, the main decay of the $\tilde\tau_1$ is into gauginos in particular into the bino-like neutralino for $\tilde{\tau}_1 \sim \tilde{\tau}_R $. For all our
benchmarks the $\tilde{b}_1$ is right-handed, $\tilde{b}_1$ will therefore also decay preferably into a bino if the
latter is lighter, otherwise decays into the higgsino  are preferred. For such decays it is crucial
to specify the exact value of the sbottom mass. Our tree-level calculation is done with a pole mass for the bottom,
$m_b=4.7$~GeV. If the decay is indeed dominated by the Yukawa coupling, we
should note that the use of a running $\tilde{q}$ mass at the scale of the decaying
particle, {\it i.e.} the sbottom mass of around 1 TeV, would be more appropriate
in order to take into account the bulk of the QCD corrections. Using the running
bottom mass brings in a relative correction
of order $2 \delta m_b/m_b \sim -72\% $. Indeed, at one-loop, we have $m_b^{\drbar}(\bar{\mu}=1{\rm TeV}) \sim
m_b^{\rm pole}
(1+a_s (\ln(m_b^2/\bar{\mu}^2)-5/3))$, $a_s=\alpha_s(\bar{\mu})/\pi$. From these observations we should expect a
strong $\tb$ (and scheme) dependence whenever the QCD corrections are large for sbottom decays and could be accounted
for by the running of $m_b$.
For the stop, the higgsino coupling is large due to the large  $m_t$,  therefore decays into
a higgsino will generally dominate. Likewise the correction driven by the running of the top mass is $2 \delta m_t/m_t
\sim -30\% $ at a scale of 1TeV (and about -37\% for a scale at 2TeV). In all cases decays into singlinos are
strongly disfavoured unless the singlino is the only kinematically accessible mode.

\subsection{Point 1}
\label{sec:sfermiondecay_point1}
\noi For this point all sfermions are at the TeV scale, in fact the stops  are even heavier with a mass around 2 TeV.
$\tilde\tau_1$ and $\tilde{t}_1$ are heavily mixed (between  LH and RH) while the $\tilde{b}_1$ is dominantly
RH. $\tb$ is rather large. Recall that this point is characterised as
$(\tilde{W}^0_3,\tilde{H}^0,\tilde{H}^0,\tilde{S}^0,\tilde{B}^0)$ and $(\tilde{W}^+,\tilde{H}^+)$ with rather large mixing between the winos and the higgsinos, see Table~\ref{tab:compo}.

\begin{table}[!hbt]
\caption{Point 1 : Partial decay widths (in GeV) of third generation sfermions into a fermion and a neutralino/chargino at
tree-level (tree) and at one-loop in three schemes (see text for their definition) including for the squarks both the electroweak and QCD effects. The total (electroweak and QCD) relative correction is indicated between round parentheses $(\;\;)$.  The relative QCD correction is  given in squared parentheses $[\;\;]$. The relative QCD correction is the same in both the  $t_{345}$ and $OS_{245h_2A_1A_2}$ scheme, see text. It is therefore not listed.}
\label{tab:sq:point1}
\begin{center}
\begin{tabular}{lllll}\hline 
&  tree & $t_{345}$& $OS_{245h_2A_1A_2}$ &$\drbar$ \\
\hline 
$\sbotl\to b\neuto$  \;\;&  0.210   & 0.058  (-72\%)   &   $-0.013$  (-106\%) [-68\%]  &0.065 (-69\%) [-68\%] \\
$\sbotl\to b\neutt$  \;\; & 0.551    &  0.164 (-70\%)        &$-0.034$(-106\%) [-75\%]& 0.165 (-70\%) [-75\%]\\
$\sbotl\to b\neuth$  \;\; & 0.408    & 0.133 (-67\%)     & $-0.018$  (-104\%) [-75\%] &0.126 (-69\%)[-75\%] \\
$\sbotl\to t\chargm$  \;\;  & 0.357   & 0.077 (-78\%)  & $  -0.044 $ (-112\%)[-74\%]  &0.088 (-75\%)[-74\%]\\
$\sbotl\to t\chargtm$  \;\; & 0.732    & 0.231 (-68\%) &$  -0.040  $  (-105\%) [-75\%] &0.222 (-70\%) [-75\%]\\\hline
$\stopl\to t \neutt $ \;\;  &   15.3 & 10.5  (-31\%)    &  10.6  (-31\%) [-34\%] &10.5 (-31\%)[-34\%]\\
$\stopl\to t \neuth $ \;\;  &  20.1   & 14.5 (-28\%)  &    14.7 (-27\%) [-28\%]& 14.3 (-29\%)[-28\%]\\
$\stopl\to b\chargt$  \;\;  &  23.4   &16.8 (-28\%)    &  16.7(-29\%) [-29\%] & {17.1 (-27\%)[-29\%]} \\\hline
$\tilde{\tau}_1\to\tau\neuto$ \;\; & 1.73   & 1.65 (-4\%) &  1.60 (-7\%) &1.62 (-6\%)\\
$\tilde{\tau}_1\to\nu_\tau\charg$ \;\;  & 3.13   & 3.01 (-4\%) &  2.93 (-6\%)& 2.98 (-5\%)\\\hline
\end{tabular}
\end{center}
\end{table}
\noi The decays of the $\tilde{b}_1$ are easy to understand. For this RH state the gauge decay would have been
into the bino-like neutralino, but this channel is kinematically closed. Decays are therefore totally triggered by
the $\tb$ enhanced Yukawa coupling into higgsino states which seep into $\neuto$ through $\tilde{H}-\tilde{W}$ mixing. Table~\ref{tab:sq:point1} shows large (negative) corrections for
sbottom decays with essentially the same corrections for all channels. The bulk of the corrections comes from the
running of the bottom mass, remember the $-72\%$ QCD correction, which our full calculations reproduces rather well. 
Once this correction is taken into account the remaining QCD correction is less than $5\%$ for all the channels and no scheme dependence is to be noticed for the QCD part of the corrections. As for the electroweak corrections, these are very small in both the  $t_{345}$ and $\drbar$ scheme, they do not exceed $7\%$. The electroweak corrections in the $OS_{245h_2A_1A_2}$ scheme are about $-34\%$ off compared to any of the other two schemes. This difference is due  to the large finite term induced by the counterterm $\delta \tb/\tb$, recall that compared to $\drbar$ we had found a
difference of about $-17\%$, see section~\ref{sec:gaugedecay_point1}. This is exactly what is needed to account for the difference between the predictions of the two schemes. 
$\Delta (\delta \Gamma/\Gamma)\simeq 2 \Delta (\delta \tb/\tb)$. We are referring to $\Delta$ as the difference between
two schemes and $\delta$ as the loop correction. This calculation shows that for such decays a very good
reconstruction (scheme) for $\tb$ is crucial.\\
\noi The decays of the $\tilde{t}_1$ proceed dominantly through the Yukawa coupling, again the bulk of the
correction is from QCD
and is accounted for by the running of the top mass, as expected. Unlike the case with $\tilde{b}_1$ the dependence on $\tb$ and therefore the scheme is hardly noticeable. The electroweak corrections here are not larger than $3\%$. 
\noi For $\tilde{\tau_1}$, the largest decays process through the gauge $SU(2)$ component, since the mixing with
the very heavy binos are unreachable and the Yukawa couplings are too small for a transition through the higgsino.  This also explain the very weak scheme dependence. 

\subsubsection{Point 2}
\label{sec:sfermiondecay_point2}
\begin{table}[!htb]
\caption{\label{tab:sq:point2} Point 2 :  Same as in Table.~\ref{tab:sq:point1}
but for Point 2. }
\begin{center}
\begin{tabular}{lllll}\hline 
& tree & $t_{123}$& $OS_{12h_2A_1A_2H^+}$  &$\drbar$ \\ \hline 
$\sbotl\to b\neutt$  \;\; &  0.332  &   0.318    (-4\%)     & 0.318  (-4\%)[-16\%]  & 0.320 (-4\%)[-16\%] \\
$\sbotl\to b\neuth$  \;\;  & 0.120   &  0.037    (-69\%)   & 0.059 (-51\%)[-72\%] &0.038 (-69\%)[-72\%] \\
$\sbotl\to b\neutf$  \;\;& 0.258   &  0.208      (-19\%) &     0.234 (-9\%)[-20\%] &0.213 (-18\%)[-20\%]  \\
$\sbotl\to t\chargm$ \;\; & 0.228 & 0.066 (-71\%) &  0.107 (-53\%)[-72\%]& 0.066 (-71\%)[-72\%] \\ \hline
$\stopl\to t \neuto$  \;\;  &  0.178  & 0.346    (94\%)     & 0.185 (4\%)[-20\%] & 0.133 (-25\%)[-20\%]\\
$\stopl\to t \neutt$  \;\;  &   0.414 &   0.241    (-42\%)   & 0.334 (-19\%) [-19\%]& 0.328 (-21\%)[-19\%]\\
$\stopl\to t \neuth$  \;\; & 0.639   & 0.572    (-11\%)       & 0.574 (-10\%)[-17\%] & 0.567 (-11\%)[-16\%] \\
$\stopl\to t \neutf$  \;\;  &  0.648  &    0.631    (-2\%) & 0.624 (-4\%)[-11\%]
& 0.648 (0\%)[-12\%] \\
$\stopl\to b \charg$ \;\; & 4.19 & 3.76 (-10\%) &  3.73 (-11\%) [-22\%]& 3.75 (-10\%)[-21\%] \\\hline
$10^{4} \times (\tilde{\tau}_1\to\tau\neuto)$ \;\;\; & 6.16 & 15.9 (141\%) & 9.27 (40\%) & 6.99 (6\%)\\\hline
\end{tabular}
\end{center}
\end{table}
As compared to Point 1,  the bino is much lighter  and at the same time the Yukawa coupling is smaller due to a smaller $\tb$ (4.5 instead of 10).  The decay of  $\tilde{b}_1 \simeq \tilde{b}_R$  is therefore dominated by the  bino when the neutralino has a fair
amount of bino.  This is the case for  $\sbotl\to b\neutt$ that is triggered by the (hypercharge) gauge coupling. Expectedly, this  decay which is not sensitive to the Yukawa of the bottom shows no scheme dependence, for both the electroweak and the QCD corrections. 
The $-16\%$ QCD correction is counterbalanced by a $ +12\%$ electroweak correction. 
When the decay is into higgsino dominated states ( $\sbotl\to t\chargm$ and $\sbotl\to b\neuth$) we reach
similar  conclusions as for  Point 1, namely the bulk of the correction is from QCD and can be
accounted for by a running of $m_b$. For  decays into higgsinos, 
the discrepancy between the $\drbar$ and $t_{123}$  schemes on the one hand and the
$OS_{12h_2A_1A_2H^+}$ scheme on the other is due to the finite part of the $\delta \tb$ contribution, see section~\ref{sec:gaugedecay_point2}. The decay into $\neutf$ involves both the bino (gauge) and the higgsino (Yukawa) couplings, the bulk of the correction is due to the running $b$ mass, while  the $10\%$ discrepancy in the electroweak corrections found for the $OS_{12h_2A_1A_2H^+}$ is due to the
$\tb$ reconstruction.

\noi Here the lightest stop is mainly $\tilde{t}_R$ and has a mass of about
$460$ GeV. Normally the dominant decays would
be to the higgsino rich states and eventually to the bino rich $\neutt$, however phase space penalises the
decays into $\stopl\to b \charg, t \neuth, t \neutf$. The corrections for these three decays are quite moderate, in
part because in the QCD corrections the running top mass should be evaluated  at lower scale and as is the case for $\stopl\to t \neutf$ the bino (gauge decay contribution) is competitive. In any case, contrary to the sbottom the $\tb$ dependence is weak as is
reflected in Table~\ref{tab:sq:point2}. Stop decays into the LSP singlino, $\neuto$,  is fraught with uncertainties. First, these decays are possible  because of the small higgsino component which through mixing allows decays to an almost singlino state. The bulk of the corrections in the fully $\drbar$ scheme is in line with a running of $m_t$ which provides about $-20\%$
corrections. The discrepancies in the other two schemes are rendered by a large correction in the finite part of
$\delta \lambda$, see the values in section~\ref{sec:gaugedecay_point2}.

\noi Due to phase space the light $\tilde{\tau}_1$ which is mostly  $\tilde{\tau}_R$  can only decay to the LSP singlino. Not
surprisingly the rate is ridiculously small. Since the only non singlino component of $\neuto$ is the higgsino, the
decay is sensitive to $\lambda \tb$, the difference between large corrections in the schemes $t_{123}$ and $OS_{12h_2A_1A_2H^+}$ on the one hand and the
fully $\drbar$ on the other hand can be explained by the finite part of the $\delta \lambda$ and $\delta \tb$ which can be found in
our discussion in section~\ref{sec:gaugedecay_point2}.

\subsubsection{Point 3}
\label{sec:sfermiondecay_point3}
\begin{table}[!htb]
\begin{center}
\caption{Point 3 : Same as in Table.~\ref{tab:sq:point1}
but for Point 3.  }
\label{tab:sq:point3}
\begin{tabular}{llllll}\hline 
& tree& $t_{134}$& $OS_{34h_2A_1A_2H^+}$&$\drbar$ \\\hline 
$\sbotl\to b\neuto$  \;\;&   0.660 & 0.180 (-73\%)    &  0.112 (-83\%)[-75\%]& 0.190 (-71\%)[-75\%]\\
$\sbotl\to b\neutt $ \;\; & 0.624 & 0.192 (-69\%) &  0.118 (-81\%) [-75\%]&0.192 (-69\%) [-75\%]\\
$\sbotl\to b\neutf$  \;\;  & 0.135   & 0.146 (8\%) &     0.146  (8\%) [-1.5\%]   &0.146 (8\%) [-1.5\%]    \\
$\sbotl\to t\chargm$ \;\; & 1.21 & 0.350 (-71\%) &  0.207 (-83\%) [-74\%]& 0.350 (-71\%)[-74\%]\\\hline
$\stopl\to t \neuto$  \;\;   & 7.59   & 5.89 (-22\%)  &   5.89 (-22\%) [-27\%]&5.88 (-23\%)[-27\%] \\
$\stopl\to t \neutt$  \;\;   &   7.89 &  5.91 (-25\%) &   5.95 (-25\%) [-26\%]& 5.93 (-25\%)[-26\%]  \\
$\stopl\to t \neutf$  \;\;  &   0.276 & 0.280 (2\%)  &   0.281 (2\%) [-0.3\%]   &0.281 (2\%)[-0.01\%] \\
$\stopl\to b \charg$ \;\; & 15.8 & 12.5 (-21\%) &  12.5 (-21\%) [-28\%]& 12.5 (-21\%)[-28\%] \\\hline
$\tilde{\tau}_1\to\tau\neuto$ \;\; & 0.116 & 0.150 (29\%) & 0.143 (23\%)&
0.120(3\%)\\
$\tilde{\tau}_1\to\tau\neutt$ \;\; & 0.0950 & 0.0820 (-14\%) & 0.0695 
(-27\%)& 0.1024 (8\%)  \\
$\tilde{\tau}_1\to\tau\neutf$ \;\; & 1.214 & 1.312 (8\%) &  1.311 (8\%) &
1.328 (9\%)\\
$\tilde{\tau}_1\to\nu_\tau\charg$ \;\; & 0.193 & 0.213 (10\%) &  0.191
(-1\%)& 0.198 (3\%) \\\hline
\end{tabular}
\end{center}
\end{table}
\noi For this point all third generation sfermions are around the TeV scale. $\tilde{\tau}_1$ is almost 
$\tilde{\tau}_R$ and a similar statement can be made for $\tilde{t}_1 \sim \tilde{t}_R$. Apart from the heaviest neutralino
and chargino which are wino-like, the other neutralinos and the chargino are kinematically accessible to all 3
sfermions studied. $\neuth$ being dominantly singlino does not show up in our list of decays, it couples to sfermions 
far too feebly. Decays to $\neuto,\neutt,\charg$ which are all higgsino-like are dominant for $\tilde{t}_1$ and $\tilde{b}_1$ and small for $\tau_1$, since these couplings are proportional to the Yukawa coupling. As for other points,  large (negative)  radiative corrections are found for these decays for both stops and especially sbottoms for all the schemes, these corrections can be incorporated in the running of $m_t$ and $m_b$.
For both $\tilde{\tau}_1$ and $\tilde{b}_1$ we notice again a non negligible scheme dependence due to the implementation of $\delta \tb$, the difference between  the $t_{123}$ and  the  $OS_{34h_2A_1A_2H^+}$    is very well accounted for  by the finite value of $\delta \tb/\tb$ of the
scheme. While for the squarks the difference between the full $\drbar$ scheme and the $t_{123}$ scheme is not noticeable, it is not
the case for the $\tilde{\tau}_1$, we have traced this difference to the implementation of the $\tilde{\tau}$ mixing angle, where we
applied different definitions for the squark and the slepton sector, see Section~\ref{sec:ren_sf}. Despite the fact that  $\neutf$ is far heavier than $\neuto$ and $\neutt$, $\tilde{\tau}_1 \to \tau \neutf$ is the largest
partial width for $\tilde\tau_1$. Decays of $\tilde{t}_1$ and especially $\tilde{b}_1$ into $\neutf$ are also not negligible. This is normal, $\neutf$ is essentially bino-like with  a relatively large coupling to $f_R$ states, in particular sleptons. In this case the radiative corrections are modest  and scheme independent. This is also not surprising since the decays are driven essentially by the $U(1)$ gauge coupling.

\subsubsection{Point 4}
\label{sec:sfermiondecay_point4}
 This  point  is the most MSSM-like with a very small $\lambda$ so that the singlino is practically decoupled. It is therefore normal that $\neuth$ does not show up in the table of decays Table~\ref{tab:sq:point4}. $\tilde{t}_1$ and $\tilde{\tau}_1$ are quite light here and are, like $\tilde{b}_1$ mainly RH. Decays to the wino dominated states $\neutt, \chi^{+}_1$ are therefore  suppressed. Because the higgsino-like state are too heavy for $\tilde{t}_1$ and $\tilde{\tau}_1$ to decay into, the only channel left for $\tilde{t}_1$ and $\tilde{\tau}_1$ is into the LSP which is  almost bino-like. 
\begin{table}[!htb]
\begin{center}
\caption{Point 4 : Same as in Table.~\ref{tab:sq:point1}
but for Point 4.}
\label{tab:sq:point4}
\begin{tabular}{lllll}\hline 
&tree & $t_{123}$& $OS_{134A_1A_2H^+}$&$\drbar$ \\\hline  
$\sbotl\to b\neuto$  \;\; &  0.508  & 0.563      (11\%)   &  0.538 (6\%) [-4\%] &0.536 (5\%)[-4\%]  \\
$\sbotl\to b\neutf$  \;\;& 0.131 &  0.030     (-75\%) &   0.101 (-23\%) [-78\%]&0.033 (-75\%) [-78\%]\\
$\sbotl\to b\neutfi$ \;\; & 0.123 & 0.016 (-87\%) &  0.094 (-23\%) [-77\%]& 0.030 (-76\%)[-77\%] \\
$10^{2}\times (\sbotl\to t\chargm)$ \;\; & $3.41$ & $0.53$ (-84\%) &
$2.04$ (-40\%) [-67\%]&$0.65$ (-81\%) [-67\%]\\
$\sbotl\to t\chargtm$ \;\; & 0.197 & 0.046 (-77\%)  & 0.151 (-23\%)[-77\%]& 0.0451 (-77\%) [-77\%]\\\hline
$\stopl\to t \neuto$  \;\;  &   0.181 & 0.210      (16\%)   & 0.211 (17\%)  [10\%] &0.21 (16\%) [10\%] \\\hline
$10^{3}\times (\tilde{\tau}_1\to\tau\neuto)$ \;\; & $8.01$ & $8.78$ (10\%) & $8.75$ (9\%) & $8.79$ (10\%) \\\hline
\end{tabular}
\end{center}
\end{table}
Radiative corrections to the decays into the LSP (bino dominated) for all three sfermions are relatively small and most importantly the 
scheme dependence is hardly noticeable. This is as expected since these transitions are triggered by the $U(1)$ gauge coupling. For the 
sbottom, the other decays heavily involve the higgsino component. Again it is the same story, the large negative QCD corrections are 
accounted for by the running of $m_b$ and the discrepancy between the full $\drbar$, the $t_{123}$ and the  $OS_{134A_1A_2H^+}$ 
scheme are accounted for by the large contribution from the finite part of the $\delta \tb$ counterterm derived in the latter scheme.

\subsubsection{Point 5}
\label{sec:sfermiondecay_point5}
\noi Compared  to the previous point, Point 4, the nature of the stop and sbottom has not changed. \\
\noi $\tilde{t}_1$ which is mainly  $
\tilde{t}_R$ is not very heavy and the would-be preferred decays into  the bino-like, $\neuth$ and higgsino-like states, $
\neutf,\neutfi,\tilde{\chi}^+_2$ are kinematically not possible. Decays into the remaining wino-like, $\neuto$ and singlino-like, $\neutt$, 
state  are  extremely suppressed as Table~\ref{tab:sq:point5} shows. Note that Point 5 has the largest value of $\lambda$ of all the 
benchmarks we proposed. Although the  amount of singlino mixing  remains small, $\tilde{t}_1$ decays  into the singlino  as it is the only kinematically accessible state in this category. 
The decay is inherited from the Yukawa higgsino coupling and transmitted then to the singlino-rich $\neutt$. With stops, the $\tb$ dependence 
 is small but the singlino parameter $\lambda$ controls this decay. It turns out that the difference between the three 
schemes is quite small and follows from the fact the finite part of $\delta \lambda$ is quite small (contrary to what  is found for many 
of  the points we studied), see Section~\ref{sec:gaugedecay_point5}. It follows also that a large part of the correction is due to QCD 
and is encoded in the running of the top mass, while the electroweak corrections amount to less than $+10\%$.\\

\begin{table}[!bht]
\begin{center}
\caption{Point 5 : Same as in Table.~\ref{tab:sq:point1}
but for Point 5. }
\label{tab:sq:point5}
\begin{tabular}{lllll}\hline 
&tree & $t_{123}$&$OS_{234A_1A_2H^+}$&$\drbar$ \\\hline 
$10^{3} \times(\stopl\to t \neutt)$ \;\; & $9.17$ & $7.71$ (-16\%) & 
$8.28$ (-10\%)[-17\%]& $8.18$ (-11\%)[-19\%]\\\hline
$\sbotl\to b\neuth$  \;\; &   0.362 & 0.383     (6\%)   &0.383 (6\%) [-3\%]&0.383 (6\%)[-3\%]\\ 
$10^{2} \times(\sbotl\to b\neutf$) \;\; & $3.88$ & $1.12$ (-71\%) &
$1.17$ (-70\%) [-74\%]& $1.12$ (-71\%)[-74\%]\\
$10^{2} \times(\sbotl\to b\neutfi)$ \;\; & $4.45$ & $2.16$ (-52\%) & 
$2.21$ (-50\%) [-57\%]& $2.15$ (-52\%) [-57\%]\\
$10^{2} \times(\sbotl\to t\chargt)$ \;\; & $6.17$ & $1.77$ (-71\%) & $1.86  (-70\%)$[-74\%] & $1.75$ (-72\%) [-74\%]\\\hline
$\tilde{\tau}_1\to\nu_\tau\charg$ \;\; & 7.07 & 6.912 (-2\%) &  6.98 (-1\%)
& 6.872 (-3\%) \\
$\tilde{\tau}_1\to\tau\neuto$ \;\; & 3.49 & 3.409 (-2\%) &  3.441  (-1\%) & 
3.381  (-3\%)\\
$\tilde{\tau}_1\to\tau\neuth$ \;\; & 1.04 & 1.129 (9\%) &   1.104 (6\%) &
1.145 (10\%)  \\
\hline
\end{tabular}
\end{center}
\end{table}

\noi $\tilde{b}_1$ is much heavier than $\tilde{t}_1$, in particular the channel into the bino-rich $\neuth$ is open. This constitutes the 
largest partial width for  $\tilde{b}_1$. Again, since this is mediated by the hypercharge gauge coupling, the corrections are modest 
and scheme independent with very small corrections for both the QCD and the electroweak part. 
The  decays of $\tilde{b}_1$ to the heavier higgsino-dominated neutralinos and charginos are 
Yukawa induced especially for the 
almost pure $\neutf,\tilde{\chi}^+_2$ states. Note also that $\neutfi$ has a non 
negligible bino component that seeps in also, see Table~\ref{tab:compo}. The large negative corrections are essentially QCD 
corrections that 
can be  accounted for by a running of the $b$ mass as we explained for similar cases before. Note that this time the 
$OS_{234A_1A_2H^+}$ does not differ by 
more that $2\%$ from the other schemes for these type of decays. This again is easily understood on the basis of the finite part for $\delta 
\tb/\tb$ that we calculated for this point, see Section~\ref{sec:gaugedecay_point5}. \\

\noi For the $\tilde{\tau}_1$ which has a large $\tilde{\tau}_L$ component, decays to the wino-like states $\neuto, \tilde{\chi}^+_1$ dominate, note the (almost) factor 2 ratio between the charged and neutral channels (due to isospin).  Decays into the bino-rich 
$\neuth$ are not small, of order $s_W^2/c_W^2$ compared to the  decays into the wino dominated $\neuto$. The $\tilde{\tau}_R$ component is not large enough to participate in the coupling. 
Since these decays are driven by couplings of a gauge origin there is very little scheme dependence.

\section{Conclusions}
\label{sec:conclusions}

The present paper is the first in a series that addresses the renormalisation, at one-loop, of the NMSSM paying particular attention to the implementation of on-shell schemes. We have concentrated here on the neutralino/chargino system and exposed the sfermion sector. We also appealed to some issues and features that reside within the Higgs sector and which help in better defining some key parameters which are also of importance when studying observables that only involve the neutralinos, charginos and sfermions. Details of the Higgs sector are left for a follow-up paper. After presenting the theoretical set-up, in particular the different schemes that allow to define the necessary counterterms for a complete renormalisation of the chargino/neutralino and sfermion systems, we turn to two classes of applications. For this, we have first defined a set of five benchmark points which select different hierarchies of neutralinos and charginos depending on the nature of these particles (wino-like, singlino-like, bino-like, higgsino-like and mixed). In the first  class of applications we studied the electroweak radiative correction for the decays of the type $\tilde{\chi} \to \tilde{\chi}^\prime V, V=W^\pm, Z, \tilde{\chi}^{(\prime)}=\tilde{\chi}^0,\tilde{\chi}^+$. In the second class we considered sfermion decays to a fermion and a chargino/neutralino ($\tilde{f} \to f^\prime \chi$), in particular we calculated the one-loop QCD and electroweak corrections to the lightest stop and sbottom and the electroweak corrections to the lightest stau. The results are presented for different on-shell renormalisation schemes and compared to a full $\drbar$ scheme. Considering the importance of the ubiquitous $\tb$, we also study a mixed scheme which is essentially OS apart from  a $\drbar$ implementation of $\tb$. All these calculations are obtained with {\tt SloopS} a code for the automatic generation of counterterms and the calculations of corrected masses, decays and cross sections. The theoretical set-up that we have detailed in this paper is now fully implemented in {\tt SloopS}. 

\noi The OS schemes  we have presented in this study are based on the use of a minimal set of physical masses with the view of reconstructing the totality of the underlying parameters of the NMSSM in order that any observable can be predicted at the one-loop order. Obviously there are different choices for the minimal set of physical masses.  One would have thought that if one is interested in the chargino/neutralino system, providing the masses for a subset of these particles should have been sufficient to determine all the needed counterterms. Algebraically this is indeed the case, however masses of the neutralinos and charginos are not very sensitive to some key parameters such as $\tb$ and $\lambda$. The latter sets  the amount of mixing between the singlet and the other (MSSM-like) components. As a consequence, when we study decays which are  much more sensitive to some of these parameters, we may end up with large radiative corrections due to  ill-reconstructed mixing parameters. We have studied how one can improve the predictions by trading off some of the neutralino masses by some Higgs masses since the Higgs sector also experiences mixing that are driven by the same parameters. Our results  indicate that a judicious choice of Higgs masses leads to significant improvement in the reconstruction of $\lambda$ while issues remain with $\tb$, even though some improvement on $\tb$ is always found. 
The conclusion is that  one should use as an input parameter an observable other than a mass, say a decay such as the decay of one of the neutral pseudo-scalar Higgses to $b \bar b$ as was suggested for the MSSM, see~\cite{Baro:2008bg}. 

\noi The electroweak corrections to the decays $\chi \to \chi^\prime V$ are generally modest, within $20\%$ and often much less. Larger corrections do show up in some schemes but these are due to a large contribution for the finite part of the counterterm of $\tb$ and/or $\lambda$ when those are extracted from a system of masses which is marginally affected by $\tb$ thus explaining   the large finite part of the counterterm. For sbottom decays the QCD one-loop calculation reveals corrections of order $-70\%$ and about $-20\%$ for some stop decays. These large corrections happen when the decays are triggered through the higgsino coupling. The large QCD corrections can be  absorbed, almost entirely, in the running of the respective quark masses, set at the scale of the sbottom/stop mass.\\
The different renormalisation schemes described here and the extension of {\tt SloopS} to include the NMSSM can now be used to compute any
scattering process, in particular processes involving dark matter particles relevant for computing the relic density or processes for sparticle production and decays needed for collider physics.

\section{Acknowledgements}
This research was supported in part by the French ANR, Project DMAstro-LHC, ANR-12-BS05-0006, by the {\it
Investissements d'avenir}, Labex ENIGMASS and by the Research Executive Agency
(REA) of the European Union under the Grant Agreement PITN-GA2012-316704
("HiggsTools"). The work of GC is supported by the Theory-LHC-France initiative
of CNRS/IN2P3.

\bibliography{nmssm_sloops}{}

\end{document}